# Microwave magnetoelectric fields

E.O. Kamenetskii

Department of Electrical and Computer Engineering,
Ben Gurion University of the Negev, Beer Sheva, Israel

November 16, 2011

**Abstract**

We show that in a source-free subwavelength region of microwave fields there can exist the field structures with local coupling between the time-varying electric and magnetic fields differing from the electric-magnetic coupling in regular-propagating free-space electromagnetic waves. To distinguish such field structures from regular electromagnetic (EM) field structures, we term them as magnetoelectric (ME) fields. We study a structure and conservation laws of microwave ME near fields. We show that there exist sources of microwave ME near fields – the ME particles. These particles are represented by small quasi-2D ferrite disks with the magnetic-dipolar-oscillation spectra. The near fields originated from such particles are characterized by topologically distinctive power-flow vortices, non-zero helicity, and a torsion degree of freedom. Our studies of the microwave ME near fields are combined in two successive papers. In this paper we give a theoretical background of properties of the electric and magnetic fields inside and outside of a ferrite particle with magnetic-dipolar-oscillation spectra resulting in appearance of the microwave ME near fields. Based on the obtained structures of the ME near fields, we discuss effects of so-called ME interactions observed in artificial electromagnetic materials. In the next paper, we represent numerical and experimental studies of the microwave ME near fields and their interactions with matter.

**I. Introduction**

The concept of ME near fields arises from an idea of local coupling between the time-varying electric and magnetic fields. The "local" means an assumption that a free-space region, inside which one can observe such a field coupling, is much less than the free-space plane-wave EM wavelength. In such an EM subwavelength region – the ME-field region – the nature of the electric-magnetic coupling is different from the nature of the electric-magnetic coupling in a regular-propagating free-space EM wave. It means that in the ME-field region, symmetry between the time-varying electric and magnetic fields (the "electric-magnetic democracy" – a nice term coined by Berry [1]) should be in a distinctive form compared to a symmetry relationship between the time-varying electric and magnetic fields in a free-space EM plane wave.

The problem of ME near fields incorporates a number of fundamental physical aspects. There are, for example, such questions as topological phases, chirality and helicity of the fields, and a torsion degree of freedom of the fields. In a point ME particle – a source of the ME fields – the coupling between the electric and magnetic fields is due to specific topological (geometric) phases. For a regular-wave EM field, the ME-field region behaves as a topological singularity. It is well known that topological phase singularities can be constructed based on an interference process of regular EM waves in free space. The connection between topological phase singularities in 3D free space and the regular-wave decomposition is complicated as the position of nodes depends nonlinearly on the amplitudes and directions of the regular waves. In particular, free-space optical vortices are created by interference of three [2] or more [3] plane waves, as well as by interference of spherical waves [4]. A combination of multiple free-space regular EM waves may result in other very specific singular properties. Recently, a superposition of multiple free-space plane waves was analyzed on the subject of near-field experiments with chiral molecule structures [5 – 7]. As one of parameters characterizing such a non-plane-wave EM field, it was used a so-called the optical chirality density (OCD). This quadratic-form quantity, appeared, for the first time, in the Lipkin's analysis [8], does not have an evident physical meaning in a view of some general discussions on conservation laws in the free-space Maxwell electrodynamics [8 – 11]. Using Gaussian units, we represent here the OCD as:

$$C = \frac{1}{8\pi}\left(\vec{E}\cdot\nabla\times\vec{E} + \vec{H}\cdot\nabla\times\vec{H}\right). \qquad (1)$$

Here $\vec{E}$ and $\vec{H}$ are, respectively, the real electric and magnetic fields. As it was discussed in Refs. [5, 6], the EM fields with non-zero OCD, can effectively interact with particles characterizing by the cross polarization effects. This assumes interaction of the electric field with electric dipole moments induced by the magnetic field and interaction of the magnetic field with magnetic dipole moment induced by the electric field. In tiny regions of space, the authors of Ref. [5] say, the "superchiral" light – the light with non-zero parameter $C$ – would twist around at rates hundreds of times higher than ordinary circularly polarized light. As it was shown in Ref. [7], the near fields of plasmon-resonance planar-chiral-metamaterial structural elements [12, 13] have non-zero chiral-density parameter $C$. Based on numerical results in Ref. [7], one can observe separate free-space regions with positive and negative quantities of parameter $C$. This allowed effective spectroscopic characterization of chiral biomolecules [7]. Nevertheless, from general electrodynamics aspects, a physical meaning of the light twistness proposed in Refs. [5, 6] remains unclear. As it was discussed in Ref. [11], the chirality density (1) is directly related to the polarization helicity of light in the momentum (free-space plane-wave) representations. This is not, however, the case of studies in Ref. [7] where the evanescent modes are taken into consideration. For near fields, the question, how parameter $C$ is related to helicity of light, is open.

Formally, the terms in Eq. (1) correspond to general definitions of helicity in hydrodynamics and in plasma physics. The density of hydrodynamical helicity is defined as $\vec{u}\cdot\nabla\times\vec{u}$, where $\vec{u}$ is the fluid velocity. In plasma physics, the density of magnetic helicity is defined as $\vec{A}\cdot\nabla\times\vec{A}$, where $\vec{A}$ is the vector potential of the magnetic field strength. As it was discussed in Ref. [14], the properties of helicity can be observed also for two circularly polarized transverse electromagnetic waves propagating in vacuum opposite each other (in such a way that their Poynting vectors are cancelled out). The authors of Ref. [14] showed that in this case one has an EM field with $\vec{E}\parallel\vec{H}$ and the magnetic helicity characterizing by non-zero quantity of $\vec{A}\cdot\nabla\times\vec{A}$. This statement, however, was disproved in Ref. [15], where it was shown that the field energy for the Chu and Ohkawa [14] solution diverges and the time average of $\vec{E}\cdot\vec{H}$ is zero. Also in Ref. [16], based on symmetry and topological arguments, it was shown that the claim by Chu and Ohkawa [14], that a general class of transverse EM waves in vacuum with the $\vec{E}\parallel\vec{H}$ field structure exists, is false. Lee [16] stated, in particular, that the possible existence of waves with the $\vec{E}\parallel\vec{H}$ field structure may require that the boundary conditions for the physical space be rather more complex and nontrivial than used in the Chu and Ohkawa analysis.

Which term should we properly use for a description of the light twistness in free space: "chirality" or "helicity"? Discussing this problem, it is relevant to concern some fundamental physical aspects on the relation between chirality and helicity. Mathematically, helicity is the sign of the projection of the spin vector onto the momentum vector. In elementary particle physics, helicity represents the projection of the particle spin at the direction of motion. Helicity is conserved for both massive and massless particles. Chirality (the property related to handedness) is the same as the helicity only when the particle mass is zero or it can be neglected. In condensed matter physics, chirality is to be associated with enantiomorphic pairs which induce optical activity. The existence of enantiomorphic pairs, in a crystalline sense, requires the lack of a center of symmetry. This is a necessary condition for optical activity. At the same time, the wave helicity, related to a Faraday effect, does not require a lack of a center of symmetry. While the rotation of the plane of polarization by optical activity is a reciprocal phenomenon, rotation of the plane of polarization by the Faraday effect is a non-reciprocal phenomenon. Regarding the results in Ref. [7], the observed effect of free-space optical chirality is related to the handedness property: the results are obtained with enantiomorphic pairs of plasmon-resonance planar-chiral-metamaterial structural elements [12, 13]. The question, however arises: Whether the observed effect of the free-space light twistness is really a near-field (or subwavelength) effect? Optical plasmonic oscillations in metallic particles are electrostatic resonances. For electrically small particles, these resonances are not accompanied with any electromagnetic retardation effects. One has the free space-retardation electromagnetic effects, when sizes of the plasmonic-resonance particle become comparable with the free-space electromagnetic wavelength. As we can see from Ref. [7], an in-plane size of the metallic gammadion structure used for creation of superchiral fields is minimum a half of the free-space electromagnetic wavelength. So the effect of optical chirality in Ref. [7] cannot be considered as a pure subwavelength effect.

With respect to electrostatic (plasmonic) resonances in small metallic particles at optical frequency, one can consider, to a certain extent, magnetostatic (MS) [or magnetic-dipolar-mode (MDM)] resonances in small ferrite samples as a dual effect at microwave frequencies [17, 18]. There is, however, a fundamental difference between these two effects. In a case of metal particles with electrostatic (plasmonic) resonances one has strong localization of *electric* fields in a subwavelength region [19, 20]. The description is based on the assumption that the velocities involved are sufficiently low so that the magnetic field can be neglected. It is evident that in experiments of Ref. [7], a role of electrostatic (plasmonic) resonances is to enhance an electric energy of EM fields. In contrast, in small ferrite particles with MDM oscillations one has strong localization of both *magnetic and electric* fields in a subwavelength region. This results in appearance of subwavelength power-flow-density vortices [17, 18, 21 – 24]. The phase coherence in magnetic dipole-dipole interactions in a qusi-2D ferrite disk is in the heart of the explanation of many interesting phenomena observed both inside a ferrite and in an exterior (vacuum) region. Because of dynamics of the magnetization motion in a ferrite disk, characterizing by azimuth symmetry breaking, small ferrite particles with MDM spectra originate free-space microwave fields with unique symmetry properties [17, 18]. We consider small quasi-2D ferrite disks with the MDM spectra as point sources of peculiar microwave fields – the ME fields. As we will show in this paper, a structure of such a ME field is characterized by helicity properties and a torsion degree of freedom.

Due to the ANSOFT HFSS program, one can well observe some of the ME-field properties in a microwave subwavelength region of regular EM waves. The HFSS program, in fact, composes the field structures from interferences of multiple plane EM waves inside and outside a ferrite particle. We showed that such a very complicated EM-wave process in a numerical representation can be well modeled analytically with use of a so-called magnetostatic-potential (MS-potential) wave function $\psi$. A boundary-value-problem solution for a scalar wave function $\psi$ in a quasi-2D ferrite disk shows a multiresonance MDM spectrum and unique topological vortex structures of the mode fields. There is a very good correspondence between the results of numerical (based on the HFSS program) and analytical (based on the $\psi$-function spectra) studies [17, 18, 22, 23]. It is worth noting that the MS-potential wave function $\psi$ is not just a formal mathematical representation. This function bears a clear physical meaning. Based on the $\psi$-function spectral solutions in a quasi-2D ferrite disk, one obtains such physically observable quantities as energy eigenstates, eigen electric moments, eigen-mode power flow densities.

ME fields are parity-violation and time-varying fields with specific space-time geometry. Interestingly, the properties of such fields can be related to a so-called torsion degree of freedom – a subject of heightened interest in modern literature of the field structures. Torsion of spacetime – coupling the time and the angular coordinates of the field – might be connected with the intrinsic angular momentum of matter. The torsion-structure fields can be created by ferromagnet structures with its intrinsic ordered spin motion. In the case of a ferromagnet, the spin motion originates from fermions ("spinor matter"). It is not possible to eliminate this motion through transition to a suitable rotating frame of reference. The spin angular momentum can be considered as the source of the fields which are inseparably coupled to the geometry of spacetime [25 – 27]. However, the effects of torsion in a gravitational context are far negligible experimentally [26, 27]. At the same time, it is shown that condensed matter systems can provide useful laboratories for the study of torsion. Solid and liquid crystals with topological defects in the continuum limit can also be described by a manifold where the curvature and torsion fields are proportional to the topological charge densities of the defects [28, 29]. In uniform plasmas one can observe a torsional Alfvén mode. There is a twisting of magnetic field lines forming a concentric flux shell [30]. One of important aspects of the torsion degree of freedom concerns a torsion contribution to helicity [31]. Such topological interpretation of helicity allows describing unique symmetry properties of the ME fields originated from a MDM ferrite-disk particle. Due to the intrinsic angular momentum of "spinor matter", a quasy-2D ferrite disk with MDM oscillations (characterized by the long-range phase coherence in magnetic dipole-dipole interactions between a pair of spins) can behave as a torsional defect for propagating-wave EM fields. Existence of a torsion degree of freedom can be considered as one of the most important distinctive features of the ME fields created by MDM particles.

Our studies of microwave ME fields originated from MDM ferrite particles are combined in two successive papers. In this paper we give theoretical background of properties of microwave ME near fields. In our forthcoming paper [32], we give the results of numerical and experimental studies of microwave ME near fields and their interactions with matter.

**II. On the electrostatic and magnetostatic resonances in small particles**

It is worth starting our studies from general aspects of quasistatic oscillations and a comparative analysis of microwave magnetic-dipolar (magnetostatic) resonances in small ferrite samples with optical plasmonic (electrostatic) resonances in small metallic particles. Based on this analysis, we show why small particles with magnetostatic resonances can exhibit ME properties (coupling between the time-varying electric and magnetic fields in a subwavelength region) and no such properties can be observed in a case of small particles with electrostatic resonances.

It is well known that in a general case of small (compared to the free-space electromagnetic-wave wavelength) samples made of media with strong temporal dispersion, the role of displacement currents in Maxwell equations can be negligibly small, so oscillating fields are the quasistationary fields [33]. For a case of plasmonic (electrostatic) resonances in small metallic particles, one neglects a magnetic displacement current and has quasistationary electric fields. A dual situation is demonstrated for magnetic-dipolar (magnetostatic) resonances in small ferrite particles, where one neglects an electric displacement current. As an appropriate approach for description of quasistatic oscillations in small particles, one can use a classical formalism where the material linear response at frequency $\omega$ is described by a local bulk dielectric function – the permittivity tensor $\vec{\vec{\varepsilon}}(\omega)$ – or by a local bulk magnetic function – the permeability tensor $\vec{\vec{\mu}}(\omega)$. With such an approach (and in neglect of a displacement current) one can introduce a notion of a scalar potential: an electrostatic potential $\varphi$ for electrostatic resonances and a magnetostatic potential $\psi$ for magnetostatic resonances. It is evident that these potentials do not have the same physical meaning as in the problems of "pure" (non-time-varying) electrostatic and magnetostatic fields [33, 34]. Because of the resonant behaviors of small dielectric or small magnetic objects [confinement phenomena plus temporal-dispersion conditions of tensors $\vec{\vec{\varepsilon}}(\omega)$ or $\vec{\vec{\mu}}(\omega)$], one has scalar *wave* functions: an electrostatic-potential wave function $\phi(\vec{r},t)$ and a magnetostatic-potential wave function $\psi(\vec{r},t)$, respectively. The main note is that since we are on a level of the continuum description of media [based on tensors $\vec{\vec{\varepsilon}}(\omega)$ or $\vec{\vec{\mu}}(\omega)$], the boundary conditions for quasistatic oscillations should be imposed on scalar wave functions $\phi(\vec{r},t)$ or $\psi(\vec{r},t)$ and their derivatives, but not on the RF functions of polarization (plasmons) or magnetization (magnons). One has to keep in mind that in phenomenological models based on the effective-medium [the $\vec{\vec{\varepsilon}}(\omega)$- or $\vec{\vec{\mu}}(\omega)$- continuum] description, no electron-motion equations and boundary conditions corresponding to these equations are used.

Fundamentally, subwavelength sizes eliminate effects of the electromagnetic retardation. When one neglects displacement currents (magnetic or electric) and considers scalar functions $\phi(\vec{r},t)$ or $\psi(\vec{r},t)$ as the wave functions, one becomes faced with important questions, whether there could be the *propagation behaviors* for the quasistatic wave processes and, if any, what is the nature of these retardation effects. In a case of electrostatic resonances, the Ampere-Maxwell law gives the presence of a curl magnetic field. With this magnetic field, however, one cannot define the power-flow density of propagating electrostatic-resonance waves. Certainly, from a classical electrodynamics point of view [34], one does not have a physical mechanism describing the effects of transformation of a curl magnetic field to a potential electric field. In like manner, one can see that in a case of magnetostatic resonances, the Faraday law gives the presence of a curl electric field. Analogously, with this electric field one cannot define the power-flow density of propagating magnetostatic-resonance waves since, from a classical electrodynamics point of view, one does not have a physical mechanism describing the effects of transformation of a curl electric field to a potential magnetic field [34]. Formally, from Maxwell equations it can be shown that in a case of electrostatic resonances, characterizing by a scalar wave function $\phi(\vec{r},t)$, the time-varying electric fields cannot at all be accompanied with the RF magnetic fields. Similarly, one can show that in a case of magnetostatic resonances, characterizing by scalar wave function $\psi(\vec{r},t)$, the time-varying magnetic fields cannot at all be accompanied with the RF electric field. This fact can be perceived, in particular, from the remarks made by McDonald [35]. In frames of the quasielectrostatic approximation, we introduce electrostatic-potential function $\phi(\vec{r},t)$ neglecting the magnetic displacement current: $\frac{\partial \vec{B}}{\partial t} = 0$. At the same time, from the Maxwell equation (the Ampere-Maxwell law), $\nabla \times \vec{H} = \frac{1}{c}\frac{\partial \vec{D}}{\partial t}$, we write

$$\nabla \times \frac{\partial \vec{H}}{\partial t} = \frac{1}{c}\frac{\partial^2 \vec{D}}{\partial t^2}. \tag{2}$$

If a sample does not posses any magnetic anisotropy, we have

$$\frac{\partial^2 \vec{D}}{\partial t^2} = 0. \tag{3}$$

Similarly, in frames of the quasimagnetostatic approximation, we introduce magnetostatic-potential function $\psi(\vec{r},t)$ neglecting the electric displacement current: $\frac{\partial \vec{D}}{\partial t} = 0$. From Maxwell equation (the Faraday law), $\nabla \times \vec{E} = -\frac{1}{c}\frac{\partial \vec{B}}{\partial t}$, we obtain

$$\nabla \times \frac{\partial \vec{E}}{\partial t} = -\frac{1}{c}\frac{\partial^2 \vec{B}}{\partial t^2}. \tag{4}$$

If a sample does not posses any dielectric anisotropy, we have

$$\frac{\partial^2 \vec{B}}{\partial t^2} = 0. \tag{5}$$

As it follows from Eqs. (3) and (5), the electric field in small resonant dielectric objects as well as the magnetic field in small resonant magnetic objects vary linearly with time. This leads, however, to arbitrary large fields at early and late times, and is excluded on physical grounds. An evident conclusion suggests itself at once: the electric (for electrostatic resonances) and magnetic (for magnetostatic resonances) fields are constant quantities. Such a conclusion contradicts the fact of temporally dispersive media and thus any resonant conditions. Another conclusion is more unexpected: for a case of electrostatic resonances the Ampere-Maxwell law is not valid and for a case of magnetostatic resonances the Faraday law is not valid. The above analysis definitely means that from classical electrodynamics, the spectral problem formulated *exceptionally* for the electrostatic-potential function $\phi(\vec{r},t)$ do not presume use of magnetic fields and, similarly, the spectral problem formulated *exceptionally* for the magnetostatic-potential function $\psi(\vec{r},t)$ do not presume use of electric fields. This statement lives open a question of the propagation-wave behaviors for the quasistatic-resonance processes.

The eigenvalue problem for electrostatic resonances in nanoparticles occurs at optical frequencies when an isotropic dielectric medium exhibits strong temporal dispersion and its real part assumes a negative value. The resonant wavelengths are determined by shapes of nanonostructures and dielectric responses of constituents [36, 37]. When the material linear response is described by a bulk dielectric scalar function $\varepsilon(\omega)$, the electrostatic resonances can be found as solutions of the equation [38]:

$$\vec{\nabla} \cdot \left(\varepsilon(\vec{r})\vec{\nabla}\phi\right) = 0. \tag{6}$$

For homogeneous negative permittivity particles ($\varepsilon_p < 0$) in a uniform transparent immersion medium ($\varepsilon_s > 0$) and with use of conventional Dirichlet-Neumann boundary conditions for electrostatic-potential function, this equation acquires a form of a linear generalized eigenvalue problem:

$$\vec{\nabla} \cdot \left( \theta(\vec{r}) \vec{\nabla} \phi \right) = s \nabla^2 \phi, \tag{7}$$

where $\theta(\vec{r})$ equals 1 inside the particle and zero outside the particle, and $s = 1/\left(1 - \varepsilon_p/\varepsilon_s\right)$. The eigenmodes (surface plasmons) are orthogonal and are assumed to be normalized as [38, 39]

$$\int \phi_q^*(\vec{r}) \nabla^2 \phi_{q'}(\vec{r}) d^3 r = \delta_{q,q'}. \tag{8}$$

It was pointed out that for electrostatic resonances in nanoparticles one has a non-Hermitian eigenvalue problem with bi-orthogonal (instead of regular-orthogonal) eigenfunctions [40]. Electrostatic (plasmonic) resonance excitations, existing for particle sizes much smaller than the free-space electromagnetic wavelength, are described by the *evanescent-wave* electrostatic-potential functions $\phi(\vec{r},t)$. No retardation effects are presumed in such a description. In optics, the above electrostatic theory applies only to nanoparticles, when retardation effects are negligible.

For a spherical nanoparticle of arbitrary radius provided that the latter is much smaller than the free-space wavelength of incident radiation, the resonance permittivity values are consistent with the classical Mie theory [20]. In an analysis of scattered electromagnetic fields, a small metal particle with ES oscillations can be treated as a point electric dipole precisely oriented in space [41, 42]. Importantly, a role of the magnetic field in plasmonic oscillations becomes appreciable when one deviates from the electrostatic approximation to the full-Maxwell-equation description. The retardation effects appear when particle sizes are comparable with the free-space electromagnetic wavelength. Corrections to electrostatic resonance modes due to retardation can be found by using series expansions of the solutions to time harmonic Maxwell equations with respect to the small ratio of the object size to the free-space wavelength. There is the electromagnetic-wave process with a coupling between the electric and magnetic fields [43]. It was shown recently that anomalous light scattering with quite unusual scattering diagrams and enhanced scattering cross sections near plasmon (polariton) resonance frequencies is non-Rayleigh scattering. The observed power-flow patterns cannot be understood within the frame of a dipole approximation and the terms of higher orders with respect to size parameter $q = 2\pi a/\lambda$ should be taken into account [44 – 46].

The eigenvalue problem for magnetostatic resonances in small ferrite particles looks quite different. At microwave frequencies, in a region of a ferromagnetic resonance, ferrodielectric materials (ferrites) are characterized by strong temporal dispersion of the magnetic susceptibility [33]. A notion of the magnetic susceptibility has a physical meaning if a size of a ferrite sample $l$ is much greater than atomic scales. When a sample size $l$ is also much greater than a characteristic scale of the exchange interaction processes, one can neglect the spatial dispersion of the magnetic susceptibility. Imposing now the upper bound for $l$ as $l \ll \dfrac{\omega}{c}$, one obtains a behavior of quasistationary fields. There are essentially magnetostatic fields, but because of the temporal dispersion of the magnetic susceptibility, the fields are time-dependent and are functions of the coordinates. For small magnetic samples with strong temporal dispersion, one has so-called non-uniform ferromagnetic (or magnetostatic) resonances characterizing by resonance values of the magnetic susceptibility [33, 47 – 50]. In such ferrite samples, the electromagnetic boundary problem cannot be formally reduced to the full-Maxwell-equation description and the spectral properties are analyzed based on the Walker equation for MS-potential wave function $\psi(\vec{r},t)$ [50]:

$$\vec{\nabla} \cdot \left( \vec{\mu} \cdot \vec{\nabla} \psi \right) = 0 \qquad (9)$$

Outside a ferrite this equation becomes the Laplace equation. A distinctive feature of MS resonances in ferrite samples (in comparison with electrostatic resonances in metal nanoparticles) is the fact that because of the bias-field induced anisotropy in a ferrite one may obtain the *real-eigenvalue spectra* for scalar wave functions. Such regular multiresonance spectra are clearly observed in microwave experiments with quasi-2D ferrite disks [51 – 55]. In solving the spectral problem, the homogeneous boundary conditions are imposed on the MS-potential function and a normal component of the magnetic flux density. In such a case one obtains a quasi-Hermitian eigenvalue problem for *propagating-wave* scalar functions $\psi(\vec{r},t)$. This presumes *non-electromagnetic* retardation effects in small ferrite samples. A formulation of quasi-Hermitian eigenvalue problem and analytical spectral solutions were shown recently for MS modes in a thin-film ferrite disk [21, 24, 56, 57].

In solving a spectral problem for MS oscillations, special aspects concern properties of the RF electric fields. It is very important that a role of the electric fields in MDM ferrite particles becomes evident when one does not deviate from the MS approximation to the full-Maxwell-equation description. The problem regarding electric fields in MS resonances acquires a peculiar meaning in a view of fundamental discussions in literature on the sources of the curl electric fields in macroscopic electrodynamics of media with magnetization oscillations. It is well known that in frames of classical electrodynamics one can assume the existence of a macroscopic magnetic-current density together with a magnetic displacement current [34]. On the other hand, it is known that in a macroscopic electrodynamics, there are two models for a magnetic dipole: the amperian-current (electric current loop) model and the magnetic-charge model [58]. The choice between these models in macroscopic electrodynamics is not so evident. In our case of MDM oscillations, a choice between these two models acquires a special meaning.

Recent studies of interactions between EM fields and small ferrite particles with MDM oscillations reveal strong localization of electromagnetic energy in microwaves. It was shown that small ferrite-disk particles with MDM spectra are characterized by the vortex behaviors [22, 23]. A quasi-2D MDM ferrite disk can be modeled as a small region of media rotating with very low group velocities [17, 18]. Scattering of the EM fields from the MDM-vortex particle is purely topological. For incident EM waves, such vortex topological singularities act as traps, providing strong subwavelength confinement of the microwave fields [17, 18]. It appears that a vortex may turn out to generate a "radius of no return", beyond which the incident EM fields falls inevitably towards the vortex singularity. In such a case, the MDM vortex becomes an EM "black hole" in microwaves. An EM "black holes" in optics are well known [59]. In such optical "black holes", a vortex flow imprints a long-ranging topological effect on incident light. According to Fermat's principle [60] light rays follow the shortest optical paths in media. EM fields near the vortex particle behave as the fields in the empty but curved space-time region [59]. For external EM fields, a small MDM-vortex particle is a singular point and so is of a great paradox. What actually occurs at such an infinitesimally-defined point? When the curvature of space-time at the singularity is infinite, the Maxwell theory does not describe analytically the physical conditions at this point. All this may presume a very interesting problem for mathematical mapping of desired distortions of space-time. Presently, the idea of studying a gravitational system by replicating certain of its aspects in a laboratory environment through other, analogous, means has gained wide popularity.

We should come back now to the McDonald's statement [35] that, formally, no RF magnetic fields are available in a case of electrostatic resonances and no RF electric fields are available in a case of magnetostatic resonances. As we discussed in this section, in particles with plasmonic oscillations one has a non-Hermitian eigenvalue problem and the retardation effects appear when particle sizes are comparable with the free-space electromagnetic wavelength. So a role of the magnetic field in plasmonic oscillations becomes appreciable only when one deviates from the electrostatic approximation to the full-Maxwell-equation description. In a case of MS resonances in small ferrite particles, situation is completely different. In these particles one has specific retardation effects. The electric fields arise from the MS-wave spectral problem solutions characterizing by symmetry breakings. This results in appearance of peculiar fields – the ME fields. The main point is that these ME fields are *eigen-mode* fields of the MDM spectra.

**III. Formal structure of electric fields in ferrite particles with magnetostatic resonances**

One of the main questions of the field structures of MS resonances is the question of the relationship between the electric and magnetic fields. As we will show, the "electric-magnetic democracy" in MDM ferrite-disk particles appears with a very specific form of the field symmetry. Such fields we will characterize as the ME fields.

Differential equations for the fields inside a small non-conducting magnetic sample with strong temporal dispersion can be obtained from a full system of Maxwell equations with a formal assumption that the permittivity of a medium is equal to zero. In a region of electromagnetic transparency of a dielectric medium with temporal dispersion of the magnetic susceptibility, an averaged density of the electromagnetic energy for harmonic fields is expressed as [33]

$$\bar{U} = \frac{1}{16\pi}\left[\varepsilon E_\alpha E_\alpha^* + \frac{\partial(\omega \mu_{\alpha\beta})}{\partial \omega} H_\alpha H_\beta^*\right], \quad (10)$$

where $\varepsilon$ is the medium permittivity and $\mu_{\alpha\beta}$ are the components of the permeability tensor $\vec{\mu}$. In an assumption of negligibly small variation of electric energy in a small sample of a medium with strong temporal dispersion of the magnetic susceptibility, one obtains three differential equations instead of the four-Maxwell-equation description of electromagnetic fields [33, 47, 48]:

$$\nabla \cdot \vec{B} = 0, \quad (11)$$

$$\vec{\nabla} \times \vec{E} = -\frac{1}{c}\frac{\partial \vec{B}}{\partial t}, \quad (12)$$

$$\vec{\nabla} \times \vec{H} = 0. \quad (13)$$

Taking into account a constitutive relation

$$\vec{B} = \vec{H} + 4\pi \vec{m}, \quad (14)$$

where $\vec{m}$ is the magnetization, one obtains from Eq. (11):

$$\vec{\nabla} \cdot \vec{H} = -4\pi \vec{\nabla} \cdot \vec{m}. \quad (15)$$

At the same time, based on Eqs. (12) (13), and (14) one has for harmonic fields (with the $e^{i\omega t}$ factor):

$$\vec{\nabla} \times \vec{\nabla} \times \vec{E} = -i\frac{4\pi}{c}\omega \vec{\nabla} \times \vec{m}. \tag{16}$$

Following the Helmholtz decomposition of any vector field into and potential and curl parts [61], we have

$$\vec{m} = \vec{m}_{pot} + \vec{m}_{curl}, \tag{17}$$

where

$$\vec{\nabla} \times \vec{m}_{pot} = 0 \quad \text{and} \quad \vec{\nabla} \cdot \vec{m}_{curl} = 0. \tag{18}$$

In "pure" magnetostatics (non-time-varying fields) of ferromagnets, the potential and curl parts of the magnetization are considered as physical notions defining, respectively, the solutions for the magnetic scalar potential ($\vec{H} = -\nabla \Phi_M$) and the vector potential ($\vec{B} = \vec{\nabla} \times \vec{A}$) [34]. From our studies of MS oscillations, it follows that the magnetization field $\vec{m}_{pot}$ is related to the potential magnetic field, while the magnetization field $\vec{m}_{curl}$ is related to the curl electric field. Formally, equations for the potential magnetic field and the curl electric field can be considered as completely separate differential equations. It turns out, however, that the magnetic and electric fields are united because of the spectral properties of MS oscillations in a ferrite sample. In frames of the MS description ($\vec{H} = -\vec{\nabla}\psi$), eigen MS-potential wave functions $\psi(\vec{r},t)$ have no solutions related separately to the potential and curl parts of RF magnetization. A total magnetization field is expressed as

$$\vec{m} = -\ddot{\chi} \cdot \vec{\nabla}\psi, \tag{19}$$

where $\ddot{\chi}$ is the susceptibility tensor [48].

In a MS-resonance ferrite sample, condition $\vec{\nabla} \cdot \vec{m} \neq 0$ presumes the presence of the magnetic charge density:

$$\rho^{(m)} \equiv -\vec{\nabla} \cdot \vec{m}. \tag{20}$$

For time varying fields, one can suppose that there exists the macroscopic magnetic current density (magnetization current) introduced analogously to the electric-polarization current density [58]:

$$\vec{j}^{(m)} = \frac{\partial \vec{m}}{\partial t}. \tag{21}$$

The magnetic current density and the magnetic charge density should satisfy the conservation law:

$$\vec{\nabla} \cdot \vec{j}^{(m)} = -\frac{\partial \rho^{(m)}}{\partial t}. \tag{22}$$

Along with the above condition $\vec{\nabla} \cdot \vec{m} \neq 0$ for the magnetic current density, the condition $\vec{\nabla} \times \vec{m} \neq 0$ presumes the presence of the electric current density in macroscopic Maxwell's equations [33, 34, 58]. In macroscopic electrodynamics, in general, a question about the models for a magnetic dipole (the amperian-current model with an electric current density component $\nabla \times \vec{m}$, and/or the model, when there is a magnetic current density component $\partial \vec{m}/\partial t$) is not so evident. It was discussed, in particular, that a choice between these models depends on the way of measurement of the force on a magnetic dipole which, in nonstationary cases, is different for two models [62]. The controversies about torque and force on a magnetic dipole [58, 62 – 65] raise important questions about properties of magnetization dynamics. In a series of recent publications, the problem about magnetic currents and electric fields induced by such currents in magnetic structures appears to be a subject for numerous discussions. It was shown in Refs. [66, 67] that moving magnetic dipoles in ferromagnetic metals can induce an electric field. In paper [68], authors raised the question: Can a "pure magnetic current" induce an electric field? They considered a situation of existence of a spin current without a charge current: Spin-up electrons move to one direction while spin-down electrons move to the opposite direction. An electric field produced by a steady state spin current ("magnetic-charge current") is described by the "Biot-Savart law" [69]. It was shown that in the ring geometry there exist persistent spin currents. There are different mechanisms that can sustain a pure persistent spin current. In the absence of a conventional electromagnetic flux through the ring, the system with inhomogeneous magnetic field can support persistent spin and charge currents. The Berry-phase currents were calculated with decoupled the orbital and spin degrees of freedom [69]. At the same time, it was shown that in a non-magnetic semiconductor ring, a spin-orbit interaction can sustain a pure spin current in the absence of the external magnetic field or a magnetic flux [70, 71]. Recently, the persistent spin current (magnetization current) carried by bosonic excitations has also been predicted in a ferromagnetic Heisenberg ring with the inhomogeneous magnetic field [72]. The persistent spin currents are exhibited as topological properties of a system. It was supposed that these currents can generate electric fields which are described by the "Biot-Savart law" [71, 72]. The electric field originated from a persistent magnetic current in ring geometry is an observable quantity due to a Berry phase. A standard Berry phase is a circuit integral of the differential phase in a parameter space [73]. The Berry's connection plays the same role as the ordinary vector potential in the theory of the Aharonov-Bohm effect. The appropriate generalization of Stoke's theorem transforms a linear integral of the Berry's connection to a surface integral of the curl of the Berry's connection. The Berry's connection is gauge dependent and nonobservable, while the integrand (the Berry's curvature) of the surface integral is observable [74]. The above small survey shows a very nontrivial character of our problem of curl electric fields created by a small ferrite disk particle with microwave magnetization oscillations.

In our analysis we will distinguish two types of electric fields, which we conventionally denote as the $\vec{\mathcal{E}}$ and $\vec{\epsilon}$ fields. For harmonic fields (with the $e^{i\omega t}$ factor) the electric field $\vec{\mathcal{E}}$ inside a ferrite is described by the Faraday law:

$$\vec{\nabla} \times \vec{\mathcal{E}} = -\frac{i}{c} \omega \vec{B} = -\frac{i}{c} \omega \vec{H} - i\frac{4\pi}{c} \omega \vec{m}. \qquad (23)$$

Assuming that a ferrite disk is characterized by isotropic dielectric properties, so that inside a ferrite

$$\vec{\nabla} \cdot \vec{\mathcal{E}} = 0, \qquad (24)$$

and taking into account Eq. (16), one obtains the Poisson equation for the electric field $\vec{\mathcal{E}}$ [22]:

$$\nabla^2 \vec{\mathcal{E}} = \omega \frac{4\pi}{c^2} \vec{j}_{eff}^{(e)} . \tag{25}$$

Here we denoted

$$\vec{j}_{eff}^{(e)} \equiv ic\vec{\nabla} \times \vec{m} \tag{26}$$

as an effective electric current density. From Eq. (23) it evidently follows that outside a ferrite sample (in a dielectric or in vacuum) one has

$$\vec{\nabla} \times \vec{\mathcal{E}} = -\frac{i}{c}\omega\vec{H} , \tag{27}$$

At the same time, from Eq. (25) one has for the outside electric field

$$\nabla^2 \vec{\mathcal{E}} = 0 . \tag{28}$$

Contrarily to a case of the full-Maxwell-equation description (giving the wave equation $\nabla^2 \vec{E} + \frac{\omega^2}{c^2}\varepsilon\vec{E} = 0$), we obtained, for time-varying fields, the Laplace-type equation for the vector field $\vec{\mathcal{E}}$. Evidently, the curl electric field $\vec{\mathcal{E}}$ does not "recognize" any electric charges. It is also evident that, by virtue of the Faraday-law equation (27), outside a ferrite the electric field $\vec{\mathcal{E}}$ is perpendicular to the magnetostatic-mode field $\vec{H}$.

Experiments show that the electric field of MDM oscillations in a ferrite disk is an observable quantity. One can excite MDM oscillations by external quasistatic RF electric fields [53] and distinguish (by the MDM spectrum transformation) the dielectric properties of a surrounding medium [55]. Experimental results shown in Ref. [55] are confirmed by numerical studies [32]. From numerical studies [17, 18, 32] it follows also that for MDMs, the electric and magnetic fields in the near-field region above or below a ferrite disk are not mutually perpendicular. This, being a violation of Eq. (27), presumes unusual properties of the near fields. As we will show in this paper, the electric near field in vacuum can be strongly different from the Faraday-law field defined by Eq. (27). This is due to the fact that both the electric and magnetic fields in the near-field region arise from specific magnetization dynamics of MDMs.

We introduce now another type of a curl electric field, which we designate as the $\vec{\epsilon}$ field. This field is described by the differential equation:

$$\vec{\nabla} \times \vec{\epsilon} = -\frac{4\pi}{c}\vec{j}^{(m)} , \tag{29}$$

where, for harmonic variables, $\vec{j}^{(m)} = i\omega\vec{m}$. Formally, the $\vec{\epsilon}$ field can be considered as a part of the $\vec{\mathcal{E}}$ field inside a ferrite region [see Eq. (23)]. Inside a ferrite sample with MS oscillations, the $\vec{\epsilon}$ and $\vec{\mathcal{E}}$ electric fields may be very slightly distinguishable quantities. Really, with use of the relation $\vec{m} = \vec{\chi} \cdot \vec{H}$ in Eq. (23), we have $\vec{\nabla} \times \vec{\mathcal{E}} = -\frac{i}{c}\omega\left(\vec{I} + 4\pi\vec{\chi}\right)\vec{H}$, where $\vec{I}$ is the unit matrix. In a normally magnetized ferrite disk, the multiresonance MDM spectra are observed in a region of a ferromagnetic resonance when for $\chi$ (a diagonal component of tensor $\vec{\vec{\chi}}$) there is a relation $4\pi\chi \gg 1$ (or for a diagonal component of the permeability tensor there is $|\mu| \gg 1$) [51 – 57]. It means that for MDM resonances one has inside a ferrite sample:

$$\vec{\nabla} \times \vec{\mathcal{E}} \approx \vec{\nabla} \times \vec{\mathcal{e}} = -i\frac{4\pi}{c}\omega\vec{m} = -\frac{4\pi}{c}\vec{j}^{(m)}. \tag{30}$$

While different current sources (the currents $\vec{j}_{eff}^{(e)} = ic\vec{\nabla} \times \vec{m}$ and $\vec{j}^{(m)} = i\omega\vec{m}$) originate slightly different quantities of curl electric fields (the $\vec{\mathcal{E}}$ and $\vec{\mathcal{e}}$ fields, respectively) inside a ferrite, these fields have very different physical nature. From Eq. (29), one sees that there is an evident duality with definition of the electric field $\vec{\mathcal{e}}$ and the macroscopic magnetic field in magnetostatic problems of classical electrodynamics [34]. Formally, we can complete this analogy by adding an equation for divergence of an electric flux density. We denote this flux density as $\vec{\mathfrak{D}}$. When one supposes that the electric field $\vec{\mathcal{e}}$ is accompanied with a certain electric flux density, one can assume the presence of certain electric charges. Since there are no real electric free charges in the MDM-problem solutions, we have

$$\nabla \cdot \vec{\mathfrak{D}} = 0. \tag{31}$$

It is worth also noting that introducing the electric flux density $\vec{\mathfrak{D}}$ in our problem does not presume the presence of the electric displacement current $\partial\vec{\mathfrak{D}}/\partial t$. In frames of the MS description of a small ferrite sample, we still neglect time variations of the electric energy in comparison with time variations of the magnetic energy, both inside and outside a ferrite.

Eqs. (29) and (31) are completely analogous to their magnetostatic counterparts in classical electrodynamics [34]. To complete our description, there must be a constitutive relation between $\vec{\mathcal{e}}$ and $\vec{\mathfrak{D}}$. For a region inside a ferrite, we write this constitutive relation in a form

$$\vec{\mathfrak{D}} = \vec{\mathcal{e}} + 4\pi\vec{\mathcal{P}}_m^{(e)}, \tag{32}$$

where we denoted $\vec{\mathcal{P}}_m^{(e)}$ as effective electric polarization originated from magnetization motion.

It is well known that there exits a relativistic effect when a moving magnetization has an associated electric polarization (see, e.g. [34]). On the other hand, from general symmetry arguments it can be shown that in some magnetic structures with symmetry breaking of magnetization distribution there could be the phenomenological coupling mechanisms between the electric polarization and magnetization. When, in particular, the magnetization breaks chiral symmetry, the system can sustain a macroscopic electric polarization [75 – 77]. For MDM oscillations, the last mechanism of coupling between the electric polarization and magnetization is the most appropriate. In Ref. [23] it was shown that in quasi-2D ferrite disks, the MDMs are characterized by symmetry breaking of the magnetization structures resulting in induced electric-polarization properties. This occurs since the magnetization of the MDM is spiraling along the disk axis. Based on Eqs. (29) and (32), one has inside a ferrite

$$\vec{\nabla} \times \vec{\mathfrak{D}} = -\frac{4\pi}{c}\vec{j}^{(m)} + 4\pi\vec{\nabla} \times \vec{\mathcal{P}}_m^{(e)}. \tag{33}$$

From Eq. (29), we can see that outside a ferrite disk we have

$$\vec{\nabla} \times \vec{\mathcal{E}} = 0 \tag{34}$$

and so the electric field $\vec{\mathcal{E}}$ is a potential field:

$$\vec{\mathcal{E}} = -\nabla \Phi^{(e)}. \tag{35}$$

Here $\Phi^{(e)}$ is introduced as an effective electric scalar potential. For an outside region we have $\vec{\nabla} \cdot \vec{\mathfrak{D}} = \vec{\nabla} \cdot \vec{\mathcal{E}} = 0$ and so

$$\nabla^2 \Phi^{(e)} = 0. \tag{36}$$

While for a region outside a ferrite sample, an electric field $\vec{\mathcal{E}}$ is a curl field, an electric field $\vec{\mathcal{E}}$ is a potential field. There is no evidence that in vacuum the electric near field $\vec{\mathcal{E}}$ should be perpendicular to a magnetic near field $\vec{H}$.

In our analysis, one can consider two limit cases: (a) $\frac{1}{4\pi c}\left|\vec{j}^{(m)}\right| \gg \left|\vec{\nabla} \times \vec{\mathcal{P}}_m^{(e)}\right|$ and (b) $\frac{1}{4\pi c}\left|\vec{j}^{(m)}\right| \ll \left|\vec{\nabla} \times \vec{\mathcal{P}}_m^{(e)}\right|$. In a limit case (a), we have from Eq. (33) for a region inside a ferrite

$$\vec{\nabla} \times \vec{\mathfrak{D}} = -\frac{4\pi}{c} \vec{j}^{(m)}. \tag{37}$$

When we represent the $\vec{\mathfrak{D}}$ field as a curl of certain vector potentials

$$\vec{\mathfrak{D}} \equiv -\vec{\nabla} \times \vec{A}_{\mathcal{E}}^{(m)}, \tag{38}$$

we have the following relation between the vector potential $\vec{A}_{\mathcal{E}}^{(m)}$ and the magnetic-current source (see Appendix A):

$$\nabla^2 \vec{A}_{\mathcal{E}}^{(m)} = -\frac{4\pi}{c} \vec{j}^{(m)}. \tag{39}$$

For a limit case (b), we have inside a ferrite

$$\nabla \cdot \vec{\mathfrak{D}} = \nabla \cdot \left(\vec{\mathcal{E}} + 4\pi \vec{\mathcal{P}}_m^{(e)}\right) = 0. \tag{40}$$

This gives an electrostatic-type Poisson equation

$$\nabla^2 \Phi^{(e)} = -4\pi \rho^{(e)}, \tag{41}$$

where $\rho^{(e)}$ is an effective electric charge density defined as

$$\rho^{(e)} = -\nabla \cdot \vec{\mathcal{P}}_m^{(e)}. \tag{42}$$

It is worth noting that since MS resonances are observed only in bounded ferrites [48], the electric field $\vec{\mathcal{E}}$ should be found as a result of a solution of an integro-differential problem. This presumes very complicate forms of constitutive relation between $\vec{\mathcal{E}}$ and $\vec{\mathfrak{D}}$. For a case (a), however, we have a simple situation when a constitutive parameter is equal to unit. This gives the boundary conditions on an interface between a ferrite and vacuum: continuity of normal and tangential components of the electric field $\vec{\mathcal{E}}$.

Together with the question on the physical observability of the fields originated from a MDM ferrite disk, the quadratic conservation laws, characterizing physical properties of the fields, should be physically observable. The above analysis of the electric fields in MS oscillations raises the problem of the ambiguous definition of the power-flow density, as well as the linear and angular momentum densities. In classical electrodynamics, the Poynting vector

$$\vec{p} = \frac{c}{8\pi} \operatorname{Re}\left(\vec{E} \times \vec{H}^*\right), \tag{43}$$

obtained from the full-Maxwell-equation representation, characterizes the power flow density for propagating waves with a mutual transformation between the curl electric and curl magnetic fields. For MS waves, described by Eqs. (11) – (13), there are no evident mechanisms for a mutual transformation between the potential magnetic and curl electric fields. This leaves open the question of physical relevance of Eq. (43) for MS modes. As it was discussed in Refs. [21, 22], a relevant equation for analytical studies of the power flow density of MS waves is

$$\vec{p} = \frac{i\omega}{16\pi}\left(\psi^* \vec{B} - \psi \vec{B}^*\right). \tag{44}$$

At the same time, in the HFSS numerical analysis (which composes the field structures from interferences of multiple plane EM waves inside and outside a ferrite particle) use of Eq. (43) gives proper solutions for the power flow density of MS modes in quasi-2D ferrite disks. Based on numerical studies, one can see the power-flow-density vortices both inside a MDM ferrite disk and in the near-field region outside the particle [17, 18]. It was shown also [22, 23] that the numerically obtained [with use of Eq. (43)] topological structures of the power-flow-density vortices are in a very good correspondence with the vortex structures derived analytically from Eq. (44).

The presence of the power-flow-density vortices should presume an angular momentum of the fields. It is well known from Maxwell's theory that electromagnetic radiation carries both energy and momentum [34]. The momentum may have both linear and angular contributions. In free space, the angular momentum density is calculated as

$$\vec{\mathfrak{I}} = \frac{1}{8\pi c} \operatorname{Re}\left[\vec{r} \times \left(\vec{E} \times \vec{H}^*\right)\right] \tag{45}$$

and is related in a simple way to the Poynting vector as

$$\vec{\mathfrak{I}} = \frac{1}{c^2} \vec{r} \times \vec{p}. \tag{46}$$

In an electromagnetically dense medium, the correct classical-electrodynamics definition of electromagnetic flux has long been controversial with the main competition between the Abraham and Minkowski forms [34]. While a time-averaged quantity of Abraham's density of momentum in an electromagnetically dense medium is expressed as

$$\vec{g}_A = \frac{1}{8\pi c}\operatorname{Re}(\vec{E}\times\vec{H}^*), \qquad (47)$$

for Minkowski's density of momentum in an electromagnetically dense isotropic and homogeneous medium with the scalar material parameters $\varepsilon$ and $\mu$, one has

$$\vec{g}_M = \frac{\varepsilon\mu}{8\pi c}\operatorname{Re}(\vec{E}\times\vec{H}^*). \qquad (48)$$

In a case of a dispersive anisotropic dense medium, the question: "What happens to the momentum of a photon when it enters a medium" becomes much more complicated. In free space, the physical quantities characterizing electromagnetic fields – energy density, momentum density, and angular momentum density – are conserved. The conservation can be expressed as a continuity equation relating a density and a flux density of the conserved quantity. One has the energy balance equation (Poynting's theorem) for the energy density and the energy flux density, the continuity equation for the momentum density and the momentum flux density (the Maxwell stress tensor), the continuity equation for the angular momentum density and the angular-momentum flux density. In free space, the field has an entire structure which substantiates observance of all of these conservation laws [34]. This is not the case for a ferrite medium. Inside a ferrite medium, local circularly rotating electromagnetic fields and local rotating magnetization are mutually stipulated and the momentum densities found from the conservation laws contain contributions from both the electromagnetic field and from the matter. Generally, the terms of mechanical variables in matter can be included in the constitutive relations for the material response. A phenomenological nature of the constitutive relations leaves little room, however, for definite physical interpretation of the above conservation laws. For a small confined ferrite structure, physical picture for momentum properties starts to be more unpredictable. Propagating electromagnetic fields in free space carry momentum and angular momentum parallel to each other due to relativity requirements for transverse waves propagating at the speed of light. The situation is different in the sub-wavelength vicinity of electromagnetic field sources, i.e. in the near-field regime. Under the influence of the material environment the electromagnetic near fields are spatially non-homogeneous and so no conservation of quantities under parallel displacement and rotation can be a priory assumed.

Magnetic-dipolar fields in small saturated ferrite samples are neither "pure" electromagnetic fields nor "pure" magnetization fields. This presumes the fact that a total angular momentum of a ferrite disk should be composed with the "field part" and the "mechanics part". For stable MDM states, these two parts should compensate one another, so that a total angular momentum should be equal to zero. The "field part" of an angular momentum is due to the presence of phase topology defects in the magnetic-dipolar wave process above. The existence of "mechanics part" is of pure magnetization-dynamics origin. The only non-zero "field part" of angular momentum which we can see in the vacuum near-field region is not an evidence for the presence of the angular momentum of an entire system. The spectral analytical and numerical studies of MDMs in a normally magnetized quasi-2D MDM ferrite disk gives evidence for discrete states of eigen oscillations with the vortex structures. There are azimuthally rotating fields with the vortices of the power flow density in a subwavelength region. For a given direction of a bias magnetic field, one can see the same directions of the vortex rotations in three regions: (a) a vacuum near-field region below a ferrite disk, (b) a region inside a ferrite disk, and (c) a vacuum near-field region above a ferrite disk [17, 18, 22, 23]. It is evident that for every of the MDM eigen states, there should be a conserved angular momentum of an entire structure.

The discussed above quadratic conservation law concerning power-flow density vortices and the angular momentum density in a MDM ferrite disk are related to another quadratic conservation law – the law characterizing the field helicity. The presence of the vortex core should definitely be associated with specific symmetry properties of the fields. Formally, we can introduce a quantity corresponding to the parameter of the field helicity based on Eq. (1) and with taking into account properties of the MS description ($\nabla \times \vec{H} = 0$). For a real electric field, we have:

$$F \equiv \frac{1}{8\pi} \vec{E} \cdot \nabla \times \vec{E} \ . \tag{49}$$

This parameter we call as the helicity density of the MDM fields. Evidently, for MDM oscillations, an equation for the helicity density does not have a symmetrical form of the optical chirality density expressed by Eq. (1). With use of the Faraday law $\left( \nabla \times \vec{E} = -\frac{1}{c} \frac{\partial \vec{B}}{\partial t} \right)$ for monochromatic fields, one obtains that $\vec{E} \cdot \vec{B} \neq 0$, when $F \neq 0$. The helicity characterizes the way in which the field lines curl themselves. As we showed above, an electric field in vacuum, outside a MDM ferrite disk, is a composition of a curl electric field $\vec{\mathcal{E}}$ and a potential field $\vec{\mathscr{e}}$. This composition shows unique helicity properties of the near fields originated from a MDM ferrite disk – the microwave ME near fields. For a vacuum near-field region, we can rewrite Eq. (49) as

$$F = \frac{1}{8\pi} \vec{\mathscr{e}} \cdot \nabla \times \vec{\mathcal{E}} \ . \tag{50}$$

As we will show below, non-zero parameter *F* defined by Eq. (50) presumes the presence of a certain geometrical-phase factor giving an additional (with respect to a dynamical phase) phase shift between the electric and magnetic fields. This property of the microwave ME near fields appears as a very important topological characteristic of the microwave ME near fields, especially in a view of recent interest in topology of electromagnetic fields [83 – 86].

Following the above general consideration of the electric fields in ferrite-disk particles with MS oscillations, we clarify now the problem based on the spectral solutions for the MS-wave function $\psi$. All the unique features of the electric fields are observable only at the MS resonances. In an assumption of separation of variables, a magnetostatic-potential (MS-potential) wave function in a ferrite disk is represented in cylindrical coordinates $z, r, \theta$ as

$$\psi(r, \theta, z) = C \xi(z) \tilde{\varphi}(r, \theta), \tag{51}$$

where $\tilde{\varphi}$ is a dimensionless membrane MS-potential wave function, $\xi(z)$ is a dimensionless amplitude factor, and *C* is a dimensional coefficient. For a membrane MS-potential wave function $\tilde{\varphi}$, the boundary condition of continuity of a radial component of the magnetic flux density on a lateral surface of a ferrite disk of radius $\mathfrak{R}$ is expressed as [21, 57]:

$$\mu\left(\frac{\partial \tilde{\varphi}}{\partial r}\right)_{r=\Re^-} - \left(\frac{\partial \tilde{\varphi}}{\partial r}\right)_{r=\Re^+} = -i\frac{\mu_a}{\Re}\left(\frac{\partial \tilde{\varphi}}{\partial \theta}\right)_{r=\Re^-}, \quad (52)$$

where $\mu$ and $\mu_a$ are, respectively, diagonal and off-diagonal components of the permeability tensor $\vec{\mu}$. The term in the right-hand side of Eq. (52) has the off-diagonal component of the permeability tensor, $\mu_a$, in the first degree. There is also the first-order derivative of function $\tilde{\varphi}$ with respect to the azimuth coordinate. It means that for the MS-potential wave solutions one can distinguish the time direction (given by the direction of the magnetization precession and correlated with a sign of $\mu_a$) and the azimuth rotation direction (given by a sign of $\partial\tilde{\varphi}/\partial\theta$). For a given sign of a parameter $\mu_a$, there are different MS-potential wave functions, $\tilde{\varphi}^{(+)}$ and $\tilde{\varphi}^{(-)}$, corresponding to the positive and negative directions of the phase variations with respect to a given direction of azimuth coordinates, when $0 \leq \theta \leq 2\pi$. There is an evidence for path dependence of the problem solutions. To bring a system to its initial state one should involve the time reversal operations [18]. When, however, only one direction of a normal bias magnetic field is given, the MS wave rotating in a certain azimuth direction (either counterclockwise or clockwise) should make two rotations around a disk axis to come back to its initial state. It means that for a given direction of a bias magnetic field, a membrane function $\tilde{\varphi}$, describing MS-wave oscillations in a quasi-2D ferrite disk, behaves as a double-valued function.

To make the MDM spectral problem analytically integrable, two approaches were suggested. These approaches, distinguishing by differential operators and boundary conditions used for solving the spectral problem, give two types of the MDM oscillation spectra in a quasi-2D ferrite disk. These two approaches are named as the *G*- and *L*-modes in the magnetic-dipolar spectra [23, 24]. The MS-potential wave function $\psi$ manifests itself in different manners for every of these types of spectra. In a case of the *G*-mode spectrum, where the physically observable quantities are energy eigenstates and eigen electric moments of a MDM ferrite disk [51 – 53], the MS-potential wave function $\psi$ appears as a Hilbert-space scalar wave function [21, 56, 57]. In a case of the *L* modes, the MS-potential wave function $\psi$ is considered as a generating function for the vector harmonics of the magnetic fields [18, 22, 23, 24]. For the classical-like *L*-mode spectrum, the physically observable quantities are the eigen-mode fields and the power flow densities. Two types of the MDM spectrum (the *G* and *L* modes) appear from the fact that the permeability tensor depends both on the frequency and on a bias magnetic field: $\vec{\mu} = \vec{\mu}(\omega, \vec{H}_0)$.

The *G* modes, describing MS oscillations with respect to a bias magnetic field $\vec{H}_0$ at a constant frequency $\omega$, can be considered as the *MS-magnon modes*. At the same time, the *L*-mode, describing MS oscillations with respect to frequency $\omega$ at a constant bias magnetic field $\vec{H}_0$, are the *MS-photon modes*. There is a univocal correspondence between the *G*-mode and *L*-mode spectrum solutions. The effect of connection between the *G*- and *L*-mode spectrum solutions (which we can call conventionally as the MS-magnon – MS-photon interaction) arises from the topological phase factor on a lateral surface of a ferrite disk.

While the *G*-mode and *L*-mode spectrum solutions represent two ways of an analytical integration of the problem, the HFSS-program solutions appear as the results of the frequency-domain numerical integration. These numerical solutions include both the dynamical and topological (geometrical-phase) effects [17, 18, 22, 23]. For a proper interpretation of the numerical-integration spectra and topological properties of the microwave ME fields, we have to dwell on the main aspects of the *L*-mode and *G*-mode spectrum solutions.

**IV. Electric fields in the *L*-mode solutions**

The *L*-mode solutions appear from an analysis of MS resonances in a helical coordinate system. As it was shown in Ref. [24], in a case of a quasi-2D ferrite disk, one has double-helix resonance solutions which can be reduced to solutions in a cylindrical coordinate system. For a ferrite disk of radius $\Re$ and thickness *d* placed in a *z*-directed bias magnetic field, the *L*-mode solutions in cylindrical coordinates are represented as [22, 23]:

$$\psi(r,\theta,z,t) = C\xi(z) J_\nu\left(\frac{\beta r}{\sqrt{-\mu}}\right) e^{-i\nu\theta} e^{i\omega t} \tag{53}$$

inside a ferrite disk ($r \leq \Re$, $-d/2 \leq z \leq d/2$) and

$$\psi(r,\theta,z,t) = C\xi(z) K_\nu(\beta r) e^{-i\nu\theta} e^{i\omega t} \tag{54}$$

outside a ferrite disk (for $r \geq \Re$, $-d/2 \leq z \leq d/2$). In these equations, $\beta$ is a wavenumber of a MS wave propagating in a ferrite along *z* axis, $\nu$ is a positive integer azimuth number, $J_\nu$ and $K_\nu$ are the Bessel functions of order $\nu$ for real and imaginary arguments. The function $\xi(z)$ is defined in a general form as $\xi(z) = A\cos\beta z + B\sin\beta z$. Eqs. (53), (54) show that the modes in a ferrite disk are MS waves standing along *z* axis and propagating along an azimuth coordinate in a certain (given by a vector of a normal bias magnetic field) azimuth direction.

One can consider a MS-potential function $\psi$ as a generating function for the vector harmonics $\vec{H}$. Based on such a MS-potential function one defines the magnetic field ($\vec{H} = -\vec{\nabla}\psi$) inside a ferrite disk as

$$H_r(r,\theta,z,t) = C\xi(z)\frac{\beta}{\sqrt{-\mu}} J'_\nu\left(\frac{\beta r}{\sqrt{-\mu}}\right) e^{-i\nu\theta} e^{i\omega t}, \tag{55}$$

$$H_\theta(r,\theta,z,t) = -iC\frac{\nu}{r}\xi(z) J_\nu\left(\frac{\beta r}{\sqrt{-\mu}}\right) e^{-i\nu\theta} e^{i\omega t}, \tag{56}$$

$$H_z(r,\theta,z,t) = C\beta\, \xi'(z) J_\nu\left(\frac{\beta r}{\sqrt{-\mu}}\right) e^{-i\nu\theta} e^{i\omega t}. \tag{57}$$

Taking into account that $\vec{m} = \ddot{\chi}\cdot\vec{H}$, where the magnetic susceptibility is expressed as [48]

$$\ddot{\chi} = \begin{bmatrix} \chi & i\chi_a & 0 \\ -i\chi_a & \chi & 0 \\ 0 & 0 & 0 \end{bmatrix}, \tag{58}$$

one obtains the components of magnetization. With known magnetization distributions for MDMs one can find eigen electric polarization $\vec{\mathcal{P}}_m^{(e)}$ and eigen magnetic current $\vec{j}^{(m)}$. The electric polarization in a MDM ferrite disk, originated from the chiral-order magnetization and defined by the relationship $\vec{\mathcal{P}}_m^{(e)} \propto \vec{m} \times (\vec{\nabla} \times \vec{m})$, was analyzed in Ref. [23]. In the present study, we will dwell, however, on a case of a prevailing role of a magnetic current $\left( \frac{1}{4\pi c} \left| \vec{j}^{(m)} \right| >> \left| \vec{\nabla} \times \vec{\mathcal{P}}_m^{(e)} \right| \right)$. As we will show, such a case well describes the numerical and experimental results of MDM oscillations in a 2D ferrite disk.

The radial and azimuth components of a magnetic current $\vec{j}^{(m)} = i\omega \vec{m}$ are the following:

$$j_r^{(m)} = iC\omega \xi(z) \left[ \chi \frac{\beta}{\sqrt{-\mu}} J_\nu' \left( \frac{\beta r}{\sqrt{-\mu}} \right) + \chi_a \frac{\nu}{r} J_\nu \left( \frac{\beta r}{\sqrt{-\mu}} \right) \right] e^{-i\nu\theta} e^{i\omega t}, \tag{59}$$

$$j_\theta^{(m)} = C\omega \xi(z) \left[ \chi_a \frac{\beta}{\sqrt{-\mu}} J_\nu' \left( \frac{\beta r}{\sqrt{-\mu}} \right) + \chi \frac{\nu}{r} J_\nu \left( \frac{\beta r}{\sqrt{-\mu}} \right) \right] e^{-i\nu\theta} e^{i\omega t}. \tag{60}$$

Let us consider circular components of a magnetic current:

$$j_\pm^{(m)} = j_r^{(m)} \pm i j_\theta^{(m)}. \tag{61}$$

Based on Eqs. (59) and (60), we obtain

$$j_\pm^{(m)} = iC\omega (\chi \pm \chi_a) \xi(z) \left[ \frac{\beta}{\sqrt{-\mu}} J_\nu' \left( \frac{\beta r}{\sqrt{-\mu}} \right) \pm \frac{\nu}{r} J_\nu \left( \frac{\beta r}{\sqrt{-\mu}} \right) \right] e^{-i\nu\theta} e^{i\omega t}. \tag{62}$$

Taking into account the known relations for Bessel functions: $J_\nu'(x) = -\frac{\nu}{x} J_\nu(x) + J_{\nu-1}(x)$ and $J_\nu'(x) = \frac{\nu}{x} J_\nu(x) - J_{\nu+1}(x)$, we obtain from Eq. (62)

$$j_+^{(m)} = iC\omega (\chi + \chi_a) \frac{\beta}{\sqrt{-\mu}} \xi(z) J_{\nu-1} \left( \frac{\beta r}{\sqrt{-\mu}} \right) e^{-i\nu\theta} e^{i\omega t} \tag{63}$$

and

$$j_-^{(m)} = -iC\omega (\chi - \chi_a) \frac{\beta}{\sqrt{-\mu}} \xi(z) J_{\nu+1} \left( \frac{\beta r}{\sqrt{-\mu}} \right) e^{-i\nu\theta} e^{i\omega t}. \tag{64}$$

The quantity $\chi + \chi_a$ has strict frequency-resonance dependence, while for the quantity $\chi - \chi_a$ no resonance occurs [48]. In a frequency region of the MDM oscillation spectra, there is $|\chi + \chi_a| >> |\chi - \chi_a|$. When we take $\nu = 1$ and restrict our analysis by a central region of a disk ($r << \Re$) [57], we have, evidently, $\left| j_+^{(m)} \right| >> \left| j_-^{(m)} \right|$. In this case

$$j_+^{(m)} \simeq iC\omega(\chi+\chi_a)\frac{\beta}{\sqrt{-\mu}}\xi(z)J_0\left(\frac{\beta r}{\sqrt{-\mu}}\right)e^{-i\theta}e^{i\omega t}. \tag{65}$$

This equation describes a vector characterizing by a circular polarization and a right-hand rotation (with respect to a bias magnetic field $\vec{H}_0$ directed along a disk axis $z$). In neglect of current $j_-^{(m)}$ (assuming that $j_-^{(m)} = 0$), we have $j_r^{(m)} = -ij_\theta^{(m)}$. This shows that a central region of a disk can be considered as a domain with a homogeneously precessing magnetic current.

Now, we write Eq. (29) for components of the electric field $\vec{\mathcal{E}}$:

$$\left(\vec{\nabla}\times\vec{\mathcal{E}}\right)_r = \frac{1}{r}\frac{\partial \mathcal{E}_z}{\partial \theta} - \frac{\partial \mathcal{E}_\theta}{\partial z} = -\frac{4\pi}{c}j_r^{(m)},$$

$$\left(\vec{\nabla}\times\vec{\mathcal{E}}\right)_\theta = \frac{\partial \mathcal{E}_r}{\partial z} - \frac{\partial \mathcal{E}_z}{\partial r} = -\frac{4\pi}{c}j_\theta^{(m)}, \tag{66}$$

$$\left(\vec{\nabla}\times\vec{\mathcal{E}}\right)_z = \frac{1}{r}\frac{\partial (r\mathcal{E}_\theta)}{\partial r} - \frac{1}{r}\frac{\partial \mathcal{E}_r}{\partial \theta} = 0.$$

Evidently, homogeneous precession of a magnetic current (with only the $\theta$ and $r$ components) in a central region of a quas-2D ferrite disk causes homogeneous rotation of the electric field $\vec{\mathcal{E}}$ in a disk plane. Because of such a homogeneous rotation of the $\vec{\mathcal{E}}$ field in a central region of a ferrite disk, we can also assume that in vacuum (above and below a ferrite and closely to the plane surfaces of a disk) there are only the $\mathcal{E}_\theta$ and $\mathcal{E}_r$ components of the field. So, for the field $\vec{\mathcal{E}}$ in a central region of a disk, both inside a disk and outside near a disk, we can write

$$\left(\vec{\nabla}\times\vec{\mathcal{E}}\right)_r \simeq \frac{\partial \mathcal{E}_\theta}{\partial z} = \frac{4\pi}{c}j_r^{(m)},$$

$$\left(\vec{\nabla}\times\vec{\mathcal{E}}\right)_\theta \simeq \frac{\partial \mathcal{E}_r}{\partial z} = -\frac{4\pi}{c}j_\theta^{(m)} \tag{67}$$

The homogeneously rotating electric field $\vec{\mathcal{E}}$ is shown in Fig. 1.

We consider now circular components of an in-plane magnetic field:

$$H_\pm = H_r \pm iH_\theta. \tag{68}$$

Similar to the above analysis of circular components of a magnetic current, we can show that for $\nu=1$ and $r \ll \Re$, there is $|H_+| \gg |H_-|$. Under this condition, we get

$$H_+ \simeq C\omega\frac{\beta}{\sqrt{-\mu}}\xi(z)J_0\left(\frac{\beta r}{\sqrt{-\mu}}\right)e^{-i\theta}e^{i\omega t}. \tag{69}$$

As we can see, the in-plane components of a magnetic field $\vec{H}$ are perpendicular to the components of a magnetic current $\vec{j}^{(m)}$. At the same time, from Eqs. (67) it follows that the in-plane components of an electric field $\vec{\mathcal{E}}$ are also perpendicular to the components of a magnetic current $\vec{j}^{(m)}$. So, in the vacuum near-field region, we have for the in-plane fields that $\vec{H} \| \vec{\mathcal{E}}$. This allows writing that for in-plane components there is $\vec{\mathcal{E}} \| (\vec{\nabla} \times \vec{\mathcal{E}})$.

Based on the above equations, we can determine the helicity parameter for the time-harmonic near fields as

$$F = \frac{1}{16\pi} \operatorname{Im} \left\{ \vec{\mathcal{E}} \cdot \left( \vec{\nabla} \times \vec{\mathcal{E}} \right)^* \right\}. \tag{70}$$

We can also calculate an angle between vectors $\vec{\mathcal{E}}$ and $\vec{\nabla} \times \vec{\mathcal{E}}$:

$$\cos \alpha = \frac{\operatorname{Im} \left\{ \vec{\mathcal{E}} \cdot \left( \vec{\nabla} \times \vec{\mathcal{E}} \right)^* \right\}}{\left| \vec{\mathcal{E}} \right| \left| \vec{\nabla} \times \vec{\mathcal{E}} \right|}. \tag{71}$$

When one moves away from a disk plane along a disk axis, one expects reduction of the helicity parameter of the near field. Such reduction should be not only due to attenuation of amplitudes of vectors $\vec{\mathcal{E}}$ and $\vec{\nabla} \times \vec{\mathcal{E}}$, but also due to variation of $\sin \alpha$. This property should give us evidence of a torsion structure of the MDM near fields. Based on numerical studies, in Ref. [32] it is shown that there are really non-zero helicity parameters $F$ in the near fields above and below a ferrite disk. It is also shown that when one moves away from a disk plane along a disk axis, there is reduction of parameter $F$.

For the main thickness mode in a ferrite disk, the function $\xi(z)$ is a symmetrical function with respect to $z$ axis [57]. This gives the symmetrical-function distribution of the field $\vec{\nabla} \times \vec{\mathcal{E}}$ with respect to $z$ axis, while the field $\vec{\mathcal{E}}$ is characterized by the antisymmetrical distribution with respect to $z$ axis. As a result, one has opposite signs for the helicity density $F$ above and below a ferrite disk. These signs will change when one reorients the direction of a normal bias field. Such a statement is clear from the following. At a bias magnetic field oriented along $+z$ direction, there is the 90° phase advance of the current $j_+^{(m)}$ with respect to the magnetic field $H_+$. When one reorients a bias magnetic field along $-z$ direction, there will be the magnetic field $H_-$ and the magnetic current $j_-^{(m)}$. Also, there will be the 90° phase delay of the current $j_-^{(m)}$ with respect to the magnetic field $H_-$. Thus, at reorientation of a bias magnetic field there is the 180° phase difference between the currents $j_+^{(m)}$ and $j_-^{(m)}$. From Eq. (64), one sees that in this case the field $\vec{\mathcal{E}}$ oppositely changes its direction. It gives an opposite sign for the helicity density $F$. Fig. 2 represents qualitative distributions of the helicity density $F$ for the fields above and below a ferrite disk for different orientations of a bias magnetic field. This distributions show that the near-field structure of the MDM electric field is characterized by the space and time symmetry breakings. However, for such near fields, there are evident properties for the $\mathcal{PT}$ invariance [18]. When one makes successively the parity ($\mathcal{P}$) and time-reversal ($\mathcal{T}$) transformations, one restores the field structure.

In analytical studies of the near-field helicity density $F$, there are no difficulties in finding $\vec{\nabla} \times \vec{\mathcal{E}}$. When a spectral problem for $L$ modes is solved, one easily obtains the magnetic field $\vec{H}$ and thus the vector $\vec{\nabla} \times \vec{\mathcal{E}}$ of MDMs in a near-field region. At the same time, a problem in finding the field $\vec{\mathcal{E}}$ is not so trivial. It can be supposed that when the magnetic current is known, there exists a direct way to find the $\vec{\mathcal{E}}$ fields based on solutions of Eqs. (29) and (39). At resonance frequencies of MDMs ($\omega = \omega_{res}$), one formally represents the solutions for the electric field $\vec{\mathcal{E}}$ as

$$\left(\vec{A}_{\mathcal{E}}^{(m)}(\vec{x})\right)_{res} = \frac{1}{c} \int \frac{\left[\vec{j}^{(m)}(\vec{x}')\right]_{res}}{|\vec{x} - \vec{x}'|} d^3 x' \tag{72}$$

and

$$\left(\vec{\mathcal{E}}(\vec{x})\right)_{res} = -\vec{\nabla} \times \left(\vec{A}_{\mathcal{E}}^{(m)}(\vec{x})\right)_{res} = -\frac{1}{c} \int \frac{\left[\vec{j}^{(m)}(\vec{x}')\right]_{res} \times (\vec{x} - \vec{x}')}{|\vec{x} - \vec{x}'|^3} d^3 x'. \tag{73}$$

In these solutions, however, we have to take into account the retardation effects arising from the fact that the magnetic-current sources are azimuthally running waves. For this reason, one has the causal Green function which means that the source-point time is always earlier than the observation-point time.

To be able to find proper solutions for the $\vec{\mathcal{E}}$ fields, let us consider now a problem in views of two observers. The first one is placed in a laboratory frame. This observer identifies the phase over-running of $2\pi\nu$ during a time period of the oscillation $T$ but is unable to differentiate the "orbital" and "spin" polarization angle changes for the $\vec{\mathcal{E}}$-field vector. The second observer is placed on a frame rotating with an angular velocity $\omega_{frame} = 2\pi\nu/T$. Contrarily to the first observer, this observer can distinguish "spin" polarization angle changes for the $\vec{\mathcal{E}}$-field vector. His view corresponds to Fig. 1 in condition that the time phase $\omega t$ is frozen. It is evident that the $\vec{\mathcal{E}}$ field vectors are mutually parallel around a closed loop in a disk plane. The failure of parallel transport around a closed loop, measured by Berry's phase, is a hallmark of intrinsic curvature. Such intrinsic curvature, in our case, is due to the double-helix loops of the MDM oscillations in a quasi-2D ferrite disk [24].

As a starting point in the studies, we will exclude the retardation effect in solutions for the $\vec{\mathcal{E}}$ fields. For this purpose we will make an analysis in a rotating frame. In this case, one has to do transformation of a magnetic susceptibility tensor from a laboratory frame to a rotating reference frame. When a frame rotates at a resonance frequency of MDM resonances ($\omega = \omega_{frame} = \omega_{res}$), Eqs. (59) and (60) should be rewritten as:

$$\left[\left(j_r^{(m)}\right)_{rot}\right]_{res} = C\omega\xi(z) \left[ (\chi_{rot})_{res} \frac{\beta}{\sqrt{-\mu}} J_\nu' \left(\frac{\beta r}{\sqrt{-\mu}}\right) + \left[(\chi_a)_{rot}\right]_{res} \frac{\nu}{r} J_\nu \left(\frac{\beta r}{\sqrt{-\mu}}\right) \right] \sin(\nu\theta), \tag{74}$$

$$\left[\left(j_\theta^{(m)}\right)_{rot}\right]_{res} = C\omega\xi(z) \left[ \left[(\chi_a)_{rot}\right]_{res} \frac{\beta}{\sqrt{-\mu}} J_\nu' \left(\frac{\beta r}{\sqrt{-\mu}}\right) + (\chi_{rot})_{res} \frac{\nu}{r} J_\nu \left(\frac{\beta r}{\sqrt{-\mu}}\right) \right] \cos(\nu\theta), \tag{75}$$

where $\left(\chi_{rot}\right)_{res}$ and $\left[\left(\chi_a\right)_{rot}\right]_{res}$ are the diagonal and off-diagonal components of the susceptibility tensors in a rotating reference frame at the frequencies of MDM resonances (see Appendix B). For known distributions of $\left[\left(j_r^{(m)}\right)_{rot}\right]_{res}$ and $\left[\left(j_\theta^{(m)}\right)_{rot}\right]_{res}$ inside a ferrite disk, one can obtain solutions for $\left[\left(\vec{A}_{\vec{\mathcal{E}}}^{(m)}(\vec{x})\right)_{rot}\right]_{res}$ and $\left[\left(\vec{\mathcal{E}}(\vec{x})\right)_{rot}\right]_{res}$ in a rotating reference frame. An inversion from the rotating frame to the laboratory frame will give us the required quantities $\left(\vec{A}_{\vec{\mathcal{E}}}^{(m)}(\vec{x})\right)_{res}$ and $\left(\vec{\mathcal{E}}(\vec{x})\right)_{res}$. The $\vec{\mathcal{E}}$ field acquires geometric (topological) phase by the MDM carrying an orbital angular momentum. Since the solutions for the $\vec{\mathcal{E}}$ fields can appear only due to integration over geometrical (topological) phases, we can conclude that the electric $\vec{\mathcal{E}}$ fields in the MDM solutions in a ferrite disk are exclusively the topological fields. We can also conclude that the helicity density $F$ of the MDMs appears because of the presence of topological fields $\vec{\mathcal{E}}$. When we exclude the retardation effects, we can consider the near-field space above and below a ferrite disk as being sliced into the plates with the same in-plane distributions of the $\vec{\mathcal{E}}$-field vectors. In this case, one observes only attenuation of amplitudes of vectors $\vec{\mathcal{E}}$ without any change of an angle between vectors $\vec{\mathcal{E}}$ and $\vec{\nabla}\times\vec{\mathcal{E}}$. When, however, the retardation effects are taken into account, one will observe also variation of an angle between vectors $\vec{\mathcal{E}}$ and $\vec{\nabla}\times\vec{\mathcal{E}}$ in the near-field region. So, when the retardation effects in solutions for the $\vec{\mathcal{E}}$ fields are taken into consideration, one can observe analytically certain torsion properties of the MDM near fields.

**V. Electric fields in the *G*-mode solutions**

In the *L*-mode representation, the MS-potential $\psi$ functions serve as generating functions and the observables are the fields of MDMs. In case of *G*-modes, where the observables are energy eigenstates of MDMs, the MS-potential wave functions behave as orthogonal quantum-like scalar wave functions.

The *G*-mode solutions are characterized by the singlevalued MS-potential membrane functions. However, to satisfy the boundary conditions for magnetic flux density on a lateral surface of a disk, one has to impose a geometrical phase factor. This phase factor appears due to singular edge wave functions [21]. For a *G*-mode membrane wave function $\tilde{\eta}$, the boundary condition on a lateral surface of a ferrite disk is the following:

$$\mu\left(\frac{\partial\tilde{\eta}}{\partial r}\right)_{r=\Re^-} - \left(\frac{\partial\tilde{\eta}}{\partial r}\right)_{r=\Re^+} = 0. \qquad (76)$$

On a lateral border of a ferrite disk, the correspondence between a double-valued membrane wave function $\tilde{\varphi}$ and a singlevalued function $\tilde{\eta}$ is expressed as: $\left(\tilde{\varphi}_\pm\right)_{r=\Re^-} = \delta_\pm\left(\tilde{\eta}\right)_{r=\Re^-}$, where $\delta_\pm \equiv f_\pm e^{-iq_\pm \theta}$ is a double-valued edge wave function on contour $\mathcal{L} = 2\pi\Re$. The azimuth number $q_\pm$ is equal to $\pm\frac{1}{2}l$, where *l* is an odd quantity (*l* = 1, 3, 5, …). For amplitudes we have $f_+ = -f_-$ and $|f_\pm| = 1$. Function $\delta_\pm$ changes its sign when $\theta$ is rotated by $2\pi$ so that $e^{-iq_\pm 2\pi} = -1$. As a result, one has the energy-eigenstate spectrum of MS-mode oscillations with topological phases accumulated by the edge wave function $\delta$. On a lateral surface of a quasi-2D ferrite disk, one can distinguish two different functions $\delta_\pm$, which are the counterclockwise and clockwise rotating-wave edge functions with respect to a membrane function $\tilde{\eta}$. A line integral around a singular contour $\mathcal{L}$: $\frac{1}{\Re} \oint_\mathcal{L} \left( i \frac{\partial \delta_\pm}{\partial \theta} \right) (\delta_\pm)^* \, d\mathcal{L} = \int_0^{2\pi} \left[ \left( i \frac{\partial \delta_\pm}{\partial \theta} \right) (\delta_\pm)^* \right]_{r=\Re} d\theta$ is an observable quantity. It follows from the fact that because of such a quantity one can restore singlevaluedness (and, therefore, Hermicity) of the $G$-mode spectral problem. Because of the existing the geometrical phase factor on a lateral boundary of a ferrite disk, $G$-modes are characterized by a pseudo-electric field (the gauge field) [21]. We will denote here this pseudo-electric field by the letter $\vec{\epsilon}$.

The geometrical phase factor in the $G$-mode solution is not single-valued under continuation around a contour $\mathcal{L}$ and can be correlated with a certain vector potential $\vec{\Lambda}_\epsilon^{(m)}$. We define a geometrical phase for a MDM as [21]

$$i\Re \int_0^{2\pi} [(\vec{\nabla}_\theta \delta_\pm)(\delta_\pm)^*]_{r=\Re} \, d\theta \equiv K \oint_\mathcal{L} \left( \vec{\Lambda}_\epsilon^{(m)} \right)_\pm \cdot d\vec{\mathcal{L}} = 2\pi q_\pm. \qquad (77)$$

where $\vec{\nabla}_\theta \delta_\pm = \frac{1}{\Re} \frac{\partial \delta_\pm}{\partial \theta} \bigg|_{r=\Re} \vec{e}_\theta$ and $\vec{e}_\theta$ is a unit vector along an azimuth coordinate. In Eq. (77), $K$ is a normalization coefficient. The physical meaning of coefficient $K$ we will discuss below. Here, it is necessary to note that in Refs. [21, 55, 90], the coefficient $K$ was conventionally taken as equal to unit. In Eq. (77) we inserted a connection which is an analogue of the Berry phase. In our case, the Berry's phase is generated from the broken dynamical symmetry. The confinement effect for magnetic-dipolar oscillations requires proper phase relationships to guarantee single-valuedness of the wave functions. To compensate for sign ambiguities and thus to make wave functions single valued we added a vector-potential-type term $\vec{\Lambda}_\epsilon^{(m)}$ (the Berry connection) to the MS-potential Hamiltonian. On a singular contour $\mathcal{L} = 2\pi\Re$, the vector potential $\vec{\Lambda}_\epsilon^{(m)}$ is related to double-valued functions. It can be observable only via the circulation integral over contour $\mathcal{L}$, not pointwise. The pseudo-electric field $\vec{\epsilon}$ can be found as

$$\vec{\epsilon}_\pm = -\vec{\nabla} \times \left( \vec{\Lambda}_\epsilon^{(m)} \right)_\pm. \qquad (78)$$

The field $\vec{\epsilon}$ is the Berry curvature. In contrast to the Berry connection $\vec{\Lambda}_\epsilon^{(m)}$, which is physical only after integrating around a closed path, the Berry curvature $\vec{\epsilon}$ is a gauge-invariant local manifestation of the geometric properties of the MS-potential wavefunctions. The corresponding flux of the gauge field $\vec{\epsilon}$ through a circle of radius $\Re$ is obtained as:

$$K \int_S \left( \vec{\epsilon} \right)_\pm \cdot d\vec{S} = K \oint_\mathcal{L} \left( \vec{\Lambda}_\epsilon^{(m)} \right)_\pm \cdot d\vec{\mathcal{L}} = K \left( \Xi^{(e)} \right)_\pm = 2\pi q_\pm, \qquad (79)$$

where $\left( \Xi^{(e)} \right)_\pm$ are quantized fluxes of pseudo-electric fields. There are the positive and negative eigenfluxes. These different-sign fluxes should be inequivalent to avoid the cancellation. It is evident that while integration of the Berry curvature over the regular-coordinate angle $\theta$ is quantized in units of $2\pi$, integration over the spin-coordinate angle $\theta'$ $\left(\theta' = \frac{1}{2}\theta\right)$ is quantized in units of $\pi$. The physical meaning of coefficient $K$ in Eqs. (77), (79) concerns the property of a flux of a pseudo-electric field. It should, definitely, be related to the notion of a magnetic current in the $G$-mode analysis. As we will show, in a case of $G$ modes, magnetic currents appear due to "surface magnetic conductance". It differs from the situation with $L$ modes where a magnetic current is the magnetization current.

It is worth noting that the pseudo-electric field (the Berry curvature) $\vec{\epsilon}$ can be characterized as the density of the pseudo-electric flux and so the quantity $\frac{\partial \vec{\epsilon}}{\partial t}$ can be considered as the density of the pseudo-electric displacement current. However, in frames of the magnetostatic description (when there is a small sample of a medium with strong temporal dispersion of the magnetic susceptibility), this pseudo-electric displacement current does not define locally the magnetic field of magnetic-dipolar modes. It also follows that in our case of magnetostatic description (and contrary to an analysis of the magnetic monopole and magnetic fluxon dynamics based on of the full-Maxwell-equation description [83, 84]), the term $\frac{\partial \vec{\Lambda}_\epsilon^{(m)}}{\partial t}$ does not define a magnetic field of the MDM. Following Eq. (77), one sees that on contour $\mathcal{L} = 2\pi\Re$ the vector potential $\vec{\Lambda}_\epsilon^{(m)}$ can be represented as a gradient of a certain scalar function [85]. It means that on a border contour there is $\nabla \times \left(\vec{\Lambda}_\epsilon^{(m)}\right)_\pm = 0$. So the pseudo-electric field $\vec{\epsilon}$ is equal to zero on contour $\mathcal{L}$. However, at any points exterior to a singular contour $\mathcal{L}$, the curl the vector potential $\vec{\Lambda}_\epsilon^{(m)}$ (and, therefore, the field $\vec{\epsilon}$) is not equal to zero. To have fields $\vec{\epsilon}$ as observable quantities in every point of a square $S$, a singular contour $\mathcal{L} = 2\pi\Re$ should be excluded from a square $S$. This is possible when one assumes that $S = \pi\left(\Re^-\right)^2$, where $\Re^- = \Re - \Delta$ with $\Delta \ll \Re$.

In further analysis of the $G$-mode solutions we will stick to a slightly different model than in Refs. [21, 55, 90]. Solutions for $G$ modes are determined by the essential boundary conditions, while for $L$-modes there are the natural boundary conditions. The difference between the essential boundary conditions and the natural boundary conditions is defined by the term $-i\mu_a\left(H_\theta\right)_{r=\Re}$, where $\left(H_\theta\right)_{r=\Re}$ is an annular magnetic field [21, 57]. This singular border term, which expresses the discontinuity of a radial component of magnetic flux density for $G$-modes, can be represented as the effective surface magnetic charge density:

$$-i\mu_a\left(H_\theta\right)_{r=\Re} \equiv 4\pi\rho_s^{(m)}. \tag{80}$$

For an annular magnetic field at a given coordinate $z$, one obtains

$$\left[\left(\vec{H}_\theta(z)\right)_\pm\right]_{r=\Re} = -C\xi(z)\vec{\nabla}_\theta\left(\tilde{\varphi}_\pm\right)_{r=\Re} = -C\xi(z)\vec{\nabla}_\theta\left[\delta_\pm\left(\tilde{\eta}\right)\right]_{r=\Re}. \tag{81}$$

Generally, the magnetostatic description presumes that any close-loop line integral of a magnetic field is zero. Since, however, a magnetic field, expressed by Eq. (81), is described by a double-valued function, a close-loop integral of this field on a border contour $\mathcal{L}$ is non-zero. The field $(H_\theta)_{r=\Re}$ is a topologically distinctive, singular magnetic field. The gradient $\vec{\nabla}_\theta \left[\delta_\pm(\tilde{\eta})\right]_{r=\Re}$ is considered as the velocity of the irrotational flow on a lateral surface of a ferrite disk. When we represent a single-valued membrane function for $G$ modes as $\tilde{\eta}(r,\theta) \equiv R(r)\chi(\theta)$, where function, $R(r)$ is described by the Bessel functions and $\chi(\theta) \sim e^{-i\nu\theta}$, $\nu = \pm 1, \pm 2, \pm 3...$, the velocity of the irrotational flow on a lateral surface of a ferrite disk is defined as [21 – 24, 94]

$$(V_\theta)_\pm = \vec{\nabla}_\theta \left[\delta_\pm(\tilde{\eta})\right]_{r=\Re} = -i \frac{(\nu + q_\pm)f_\pm}{\Re} e^{-i(\nu+q_\pm)\theta} \vec{e}_\theta, \tag{82}$$

where $\vec{e}_\theta$ is the unit azimuth vector. Evidently, $\vec{\nabla} \times \vec{V}_\theta = 0$, but the circulation of the velocity $V_\theta$ around a closed contour $\mathcal{L}$ is a constant quantity. From Eqs. (80), (81), one has for the surface magnetic charge density:

$$\rho_s^{(m)} = \frac{i}{4\pi} C\mu_a \xi(z) \vec{\nabla}_\theta \left[\delta_\pm(\tilde{\eta})\right]_{r=\Re} = \frac{1}{4\pi} C\mu_a \xi(z) R_{r=\Re} \frac{(\nu + q_\pm)f_\pm}{\Re} e^{-i(\nu+q_\pm)\theta}. \tag{83}$$

By multiplying the right-hand side of Eq. (83) with $e^{i\omega t}$ and taking a real part, one obtains the real-time azimuth wave of a surface magnetic charge density.

In our problem under consideration, the phase-derivative effect of dipole-dipole interactions removes the rotational symmetry of the magnetic collective oscillations on a border ring of a ferrite disk. It is well known that in different structures of low-dimensional solids, which are described by the complex order parameter, the phase derivative effects may play an essential role. In linear-chain conductors, the time and spatial derivatives of the phase of the complex order parameter can be related to the electron charge-density waves and the electric current waves [95]. In quasi-one-dimensional metals, there are also so-called spin-density waves [96]. All these fluctuations are due to broken-symmetry ground states in metals. There are the ground states of the coherent superposition of pairs (pairs of electrons or pairs of electrons and holes) [95, 96]. It was shown that the macroscopic effect of the electric charge density waves (conductivity oscillations) is also possible in the Aharonov-Bohm-configuration structures due to the nontrivial real-space topology [97, 98]. Magnons are also the collective excitations of the ground state. In absence of spin-orbit and dipole-dipole interactions, the spin degrees of freedom are characterized by full rotational symmetry. This leads to excitations of a 1D Heisenberg antiferromagnet. Such excitations, considered as persistent magnetization currents around mesoscopic Heisenberg rings, were analyzed in Refs. [72]. In our case, one has the broken-symmetry states on a ferrite disk surface with the low-dimensional dipole-dipole magnetic-condensate waves. The magnetic charge density wave appears when the correlation length exceeds the circumference of the border ring of a ferrite disk.

For time varying $G$-mode fields, the azimuth waves of the surface magnetic charge density $\rho_s^{(m)}$, excited due to discontinuity of a normal component of the magnetic flux density, presume the presence of waves of the surface magnetic current density $\vec{j}_s^{(m)}$. These are the circulating magnetic current density waves. Both quantities, $\rho_s^{(m)}$ and $\vec{j}_s^{(m)}$, have time- and space-dependent phases. With use of separation of variables on a cylindrical surface of a ferrite disk, one has the continuity equation for the monochromatic wave process ($\sim e^{i\omega t}$):

$$\vec{\nabla}_\theta \cdot \left(\vec{j}_s^{(m)}\right)_\theta + \vec{\nabla}_z \cdot \left(\vec{j}_s^{(m)}\right)_z = -i\omega \rho_s^{(m)}, \tag{84}$$

where $\vec{\nabla}_\theta \cdot \left(\vec{j}_s^{(m)}\right)_\theta = \frac{1}{\Re} \frac{\partial \left(\vec{j}_s^{(m)}\right)_\theta}{\partial \theta}$ and $\vec{\nabla}_z \cdot \left(\vec{j}_s^{(m)}\right)_z = \frac{\partial \left(\vec{j}_s^{(m)}\right)_z}{\partial z}$. An azimuth component of the surface magnetic current density is an azimuth wave, which, at a given coordinate $z$, can be represented as

$$\left(\vec{j}_s^{(m)}(z)\right)_\theta = \left(J_s^{(m)}(z)\right)_\theta e^{-i(\nu+q_\pm)\theta} e^{i\omega t}, \tag{85}$$

where $\left(J_s^{(m)}(z)\right)_\theta$ is an amplitude. Since function $\xi(z)$ is a smooth function with a very small variation on the thickness distance of a ferrite disk [57] we assume here, as necessary approximation, that $\vec{\nabla}_z \cdot \left(\vec{j}_s^{(m)}\right)_z = 0$. With this assumption, we obtain from the above equations:

$$\left[\left(J_s^{(m)}(z)\right)_\theta\right]_\pm = \frac{1}{4\pi} C \omega \mu_a \xi(z) R_{r=\Re} f_\pm. \tag{86}$$

Let us formally associate Eq. (83) for the surface magnetic charge density and Eqs. (85), (86) for the surface magnetic current density. We can write

$$\left(\vec{j}_s^{(m)}\right)_\theta = \rho_s^{(m)} \left(\mathcal{V}^{(m)}\right)_\pm, \tag{87}$$

where

$$\left(\mathcal{V}^{(m)}\right)_\pm \equiv \frac{\omega \Re}{\nu + q_\pm} \tag{88}$$

is a certain velocity. Eq. (87) shows that the magnetic charge density wave slides along a border contour at a constant "drift" velocity $\mathcal{V}^{(m)}$. In fact, this is the phase velocity for the magnetic-charge-density wave along a border contour $\mathcal{L} = 2\pi\Re$. The velocities $\mathcal{V}^{(m)}$ are different for the positive (with the $q_+$ wavenumber) and negative (with the $q_-$ wavenumber) singular edge wave functions. It is worth noting that for a circulating magnetic current density wave, a magnetic moment (the magnetization vector) on contour $\mathcal{L}$ feels no force and undergoes the Aharonov-Bohm-type interference effect.

Similar to the vector potential $\vec{\Lambda}_\epsilon^{(m)}$ on a singular contour $\mathcal{L} = 2\pi\Re$, the surface magnetic current density $\vec{j}_s^{(m)}$ is related to double-valued functions and so can be observable only via the circulation integral over contour $\mathcal{L}$, not pointwise. Based on the integral relations, one finds solutions for a vector potential $\vec{\Lambda}_\epsilon^{(m)}$ and an electric field $\vec{\epsilon}$ in points exterior to a singular contour $\mathcal{L}$. A region of a source is an infinitesimally thin cylinder: delta-function magnetic-current loops of radius $\Re$ being summed up in a ferrite-disk height $d$. Since $\vec{j}_s^{(m)}$ is a circulating magnetic current density wave with a time-dependent phase, in these solutions we have to take into account the retardation effects. For this reason (with certain similarity to the $L$-mode solutions), one has the causal Green function which means that the source-point time is always earlier than the observation-point time.

Non-zero circulation of the velocity $V_\theta$ around a closed contour $\mathcal{L}$ results in angular momentum of the *G* mode. There is an electric (anapole) moment of the *G* mode originated from this angular momentum. The anapole moment determines an interaction of the *G* mode with an external electric field [21, 53 – 55, 90]. To find the anapole moment we introduce the following integral quantity:

$$a_{\pm}^{(e)} \equiv 4\pi\Re \int_0^d \oint_{\mathcal{L}} \left(\vec{j}_s^{(m)}(z)\right)_\theta \cdot d\vec{\mathcal{L}} dz, \tag{89}$$

The integrand for this quantity is defined as $4\pi\Re\vec{e}_r \times \left(\vec{j}_s^{(m)}(z)\right)_\theta$, where $\vec{e}_r$ is a unit vector along a disk radius. With use of Eqs. (85) and (86), Eq. (89) is written as

$$a_{\pm}^{(e)} = C\omega\mu_a\Re^2 R_{r=\Re} f_{\pm} e^{i\omega t} \int_0^d \xi(z)dz \int_0^{2\pi} e^{-i(\nu+q_{\pm})\theta} d\theta = -i2C\frac{f_{\pm}}{\nu+q_{\pm}}\omega\mu_a\Re^2 R_{r=\Re} e^{i\omega t} \int_0^d \xi(z)dz. \tag{90}$$

At the time reversal we change a sign of coefficient *f* (we have $f_+ = -f_-$). Also the time reversal changes a sign of an imaginary unit. If we assume that the azimuth numbers are $\nu = +1$ and $q_{\pm} = \pm\frac{1}{2}$, we can see that in this case vector $\vec{a}^{(e)}$ does not change its sign at the time reversal. It means that $\vec{a}^{(e)}$ is a polar vector.

Let us rewrite Eq. (90) as follows

$$a_{\pm}^{(e)} = \frac{\pi\left(I_s^{(m)}\right)_{\pm}\Re^2}{c}, \tag{91}$$

where

$$\left(I_s^{(m)}\right)_{\pm} \equiv -i\frac{2}{\pi}c\frac{f_{\pm}}{\nu+q_{\pm}}C\omega\mu_a R_{r=\Re} e^{i\omega t} \int_0^d \xi(z)dz. \tag{92}$$

One can see that the formally introduced quantity $a_{\pm}^{(e)}$ has physical meaning of the electric moment originated from a loop of the azimuthally averaged magnetic-charge current $\left(I_s^{(m)}\right)_{\pm}$. There is a parity-odd toroidal (or anapole) moment [99, 100]. One can interpret Eq. (91) as an expression which describes the electric (anapole) moment of a ferrite disk far away from the magnetic current loop ($\rho \gg \Re$) [21]. By analogy with the magnetic field originated from a loop electric current [34], one can define the electric field $\vec{\epsilon}$ in spherical coordinates ($\rho,\vartheta,\theta$) far away from a ferrite disk ($\rho \gg \Re$):

$$\left(\vec{\epsilon}_\rho\right)_{\pm} = 2\left(\frac{\pi\left(I_s^{(m)}\right)_{\pm}\Re^2}{c}\right)\frac{\cos\vartheta}{\rho^3}, \tag{93}$$

$$\left(\vec{\epsilon}_{\vartheta}\right)_{\pm} = \left(\frac{\pi\left(I_s^{(m)}\right)_{\pm}\mathfrak{R}^2}{c}\right)\frac{\sin\vartheta}{\rho^3}, \tag{94}$$

where $\vartheta$ is an inclination angle. One can see that no azimuth variation of the $\vec{\epsilon}$ field originated from magnetic-charge current $\left(I_s^{(m)}\right)_{\pm}$ is assumed in Eqs. (93), (94). This can be correct when one neglects the retardation effects in solutions for a vector potential $\vec{\Lambda}_{\epsilon}^{(m)}$ and an electric field $\vec{\epsilon}$. For a given loop current $\left(I_s^{(m)}\right)_{\pm}$, and in neglect of the retardation effects, one can formally make use of the orthogonal and complete-set vector spherical harmonics which are utilized for the magnetostatic problem solutions [34]. It is necessary to note that frequency $\omega$ corresponds to a resonance frequency of a certain MDM in a ferrite disk. So, there are spectra of azimuthally averaged magnetic-charge currents $\left(I_s^{(m)}\right)_{\pm}$ and spectra of electric moments $a_{\pm}^{(e)}$. It is worth noting also that the electric field for a "pure" (with two real point electric charges separated at a certain distance) point-like electric dipole moment $\vec{p}^{(e)}$ looks very similar to Eqs. (93), (94). In spherical coordinates ($\rho, \vartheta, \theta$) far away from a dipole, one has well known equations for the electric field [34]:

$$E_{\rho} = 2p^{(e)}\frac{\cos\vartheta}{\rho^3},$$

$$E_{\vartheta} = p^{(e)}\frac{\sin\vartheta}{\rho^3}, \tag{95}$$

$$E_{\theta} = 0$$

There is, however, a fundamental difference between the electric moment $a_{\pm}^{(e)}$ and the electric dipole moment $p^{(e)}$. Consequently, there is a fundamental difference between the electric field $\vec{\epsilon}$ defined by Eqs. (93), (94) and the electric field $\vec{E}$ defined by Eq. (95). It is evident that there is no scalar electric potential used for representing the $\vec{\epsilon}$-type electric field.

One of the main remarks of the *G*-mode analysis concerns the flux of a pseudo-electric field. It is evident that the Berry connection shown in Eq. (77) should be extended for an entire MS-potential function of a MDM, the function $\psi$. From the above equations, this function is represented as:

$$\left(\psi_{\pm}\right) = C\xi(z)\left(\tilde{\varphi}_{\pm}\right) = C\xi(z)\delta_{\pm}\tilde{\eta} = C\xi(z)\delta_{\pm}R(r)\chi(\theta), \tag{96}$$

where functions $\delta_{\pm}$ are defined on a singular contour $\mathcal{L} = 2\pi\mathfrak{R}$ and function $\tilde{\eta}$ is defined on a region $S = \pi\left(\mathfrak{R}^-\right)^2$. In connection with the function $\psi$, a total flux of the pseudo-electric field originated from a MDM ferrite disk should be written as:

$$\left(\Phi^{(e)}\right)_{\pm} = C^2 \left(\Xi^{(e)}\right)_{\pm} \int_S \tilde{\eta}\tilde{\eta}^* dS \int_0^d \xi(z)dz, \tag{97}$$

where $\int_S \tilde{\eta}\tilde{\eta}^* dS$ is a dimensionless norm of a certain $G$ mode. The flux $\left(\Phi^{(e)}\right)_{\pm}$ through the ring $\mathcal{L} = 2\pi\Re$ should be evaluated modulo the elementary flux of the electric field $\Phi_0^{(e)}$.

It is a property of a surface magnetic current $\vec{j}_s^{(m)}$ that the electric flux passing through an area bounded by such a circulating current is a quantized quantity. This quantization occurs because the MS-potential wave function must be single valued: its phase difference around a closed loop must be an integer multiple of $2\pi$. We may predict that there exists a quantum of an electric flux which should be a physical constant. A total electric flux passing through a bounded area must be a multiple of a quantum of an electric flux. However, in itself definition and, moreover, the value of a quantum of an electric flux are under a question. As one of the versions, we can represent the elementary flux of the electric field as the quantity $\Phi_0^{(e)} = 4\pi e$, where $e$ is the electron charge. It thus appears that we have

$$\frac{1}{4\pi e}\left(\Phi^{(e)}\right)_{\pm} = 2q_{\pm}. \tag{98}$$

Based on Eqs. (97), (98), one defines the normalization coefficient $K$ in Eqs. (77), (79) as follows:

$$K = \frac{C^2}{4e} \int_S \tilde{\eta}\tilde{\eta}^* dS \int_0^d \xi(z)dz. \tag{99}$$

It means that spectral properties of the MDM ferrite disk are quantized with respect to elementary electric charge. This resembles the Dirac quantization conditions. Dirac's proposition of a magnetic monopole appears from an idea of quantization of a magnetic flux. In our case, we have quantization of an electric flux.

Use of the quantity $\Phi_0^{(e)} = 4\pi e$ as an elementary flux of the electric field represents, however, a much disputed problem. In our case we have quantization of a circular magnetic current, not the magnetic charge. So an electric flux should be represented by a 2-form and not, like the electric charge, by a 3-form. Accordingly, there should be essential differences between conservation laws for the 2-form and 3-form quantities. This expresses the peculiarities of the magnetoelectric-field phenomena. At the external microwave electric field, some of the electric field lines may penetrate the ferrite in the form of thin threads of material that have turned normal. These threads, which we can call "electric fluxons" because they carry an electric flux, are in fact the central regions ("cores") of vortices in the magnetic currents. Each electric fluxon carries an integer number of electric flux quanta. The external electric field directly changes the phase of an MS-potential wave function, and it is these changes in phase that lead to measurable quantities.

For electron wave functions, the Aharonov-Bohm principle tells us that the Hamiltonian describing the system is gauge invariant under a magnetic flux changes by integral multiple of $\Phi_0^{(m)} = hc/e$, the elementary quantum of magnetic flux. Therefore, an adiabatic increase of $\Phi^{(m)}$ by a single flux quantum is a cycle of the pump in a looped ribbon. For the $G$-mode magnetostatic wave functions, the system is gauge invariant under an electric flux changes by integral multiple of the elementary quantum of an electric flux. An adiabatic increase of $\Phi^{(e)}$ by a single flux quantum is a cycle of the pump in a looped ribbon – a lateral surface of a ferrite disk.

## VI. Discussions

*A. Connection between the L- and G-mode spectra. A torsion degree of freedom for ME fields*

Two types of the MDM spectrum (the *L* and *G* modes) arise from the fact that the permeability tensor depends both on a frequency and on a bias magnetic field: $\vec{\mu} = \vec{\mu}(\omega, \vec{H}_0)$. For both these types of the spectrum solutions, the resonances take place at certain quantized states of the permeability tensor: $\vec{\mu}(\omega_{res})|_{H_0=const}$ for *L* modes and $\vec{\mu}\left[(H_0)_{res}\right]|_{\omega=const}$ for *G* modes. Since components of the permeability tensor depend non-linearly, both on a frequency and on a bias magnetic field [48], no linear correspondence exists between the resonances of the *L*- and *G*-mode spectral solutions. The *L*- and *G*- mode spectra appear with different physical properties. The *G* modes are characterized by hermitian differential operator $\hat{G}$. These modes are described by the complete-set orthogonal MS-wave functions with the energy eigenstates [21, 57]. For *L* modes one has pseudo-hermitian differential operator $\hat{L}$. The modes are described by the quasi-orthogonal MS-wave functions and are characterized by the $\mathcal{PT}$-invariance properties [18, 24]. As it was shown in Ref. [18], there exists a certain operator $\hat{\mathcal{C}}$ which provides us with connection between the *L*- and *G*-mode spectra.

The connection between the *L*- and *G*-mode spectra may manifest a certain contribution to a torsion degree of freedom for MDM oscillations, and so for ME fields. This concerns an additional spin precession which can be considered in two aspects. Firstly, it is related to the presence of the "spin-orbit" interaction term: $\vec{\epsilon}_{\pm} \times \vec{m}$. This means coupling between the gauge field $\vec{\epsilon}_{\pm}$ and magnetization $\vec{m}$. Due to the term $\vec{\epsilon}_{\pm} \times \vec{m}$ one has an interaction between a linear magnetic current $\vec{j}^{(m)}$ (*L* modes) and a circular surface magnetic current $\left(\vec{j}_s^{(m)}\right)_{\pm}$ (*G* modes). Such an interaction results in the torsion degree of freedom of the near fields. There should be two vector quantities: the vector $\vec{\epsilon}_{+} \times \vec{m}$ directed along the $+z$ axis and the vector $\vec{\epsilon}_{-} \times \vec{m}$ directed along the $-z$ axis. These two vectors give two different types of the near-field torsion structures above and below the ferrite disk. Secondly, one can observe the gravitomagnetic effect of the rotating fields [101]. For the *G*-mode spectrum, one has an anapole-moment "spin" $\vec{s}^{(e)}$ which is an intrinsic "spin" of the MDM disk particle [21, 90]. At the same time, for the *L*-mode spectrum, we have a rotating magnetic assembly. With respect to the laboratory frame, the *L*-mode fields rotate at the RF frequency $\omega$. As measured by the laboratory-frame observer the "spin" $\vec{s}^{(e)}$ must "precess" in a sense opposite to the sense of rotation of the *L*-mode fields. The Hamiltonian associated with such motion would be of the form $\mathcal{H} = -\vec{s}^{(e)} \cdot \vec{\omega}$. The existence of such a Hamiltonian would show that the intrinsic "spin" has rotational inertia. Such a gravitomagnetic effect of the rotating MDM fields can be observable only via the circulation integral over contour $\mathcal{L} = 2\pi\Re$, not pointwise. So there should be non-zero overlapping integral of double-valued and single-valued functions along contour $\mathcal{L}$. This overlapping integral is expressed as $\frac{1}{2}\int_0^{2\pi}\left[\tilde{\varphi}(\theta)\delta_{\pm}^*(\theta) - \tilde{\varphi}^*(\theta)\delta_{\pm}(\theta)\right]d\theta$, where asterisk means complex conjugation in frequency domain. It is evident that integrals $\frac{1}{2}\int_0^{2\pi}\left[\tilde{\varphi}(\theta)\delta_+^*(\theta)-\tilde{\varphi}^*(\theta)\delta_+(\theta)\right]d\theta$ and $\frac{1}{2}\int_0^{2\pi}\left[\tilde{\varphi}(\theta)\delta_-^*(\theta)-\tilde{\varphi}^*(\theta)\delta_-(\theta)\right]d\theta$ are different quantities.

*C. Interaction of ME fields with dielectrics and biological-type samples*

Microwave ME fields originated from ferrite particles with MDM oscillations are very sensitive to dielectric parameters of materials. Because of special symmetry properties, these fields should be also sensitive to a topological structure of some chemical and biological objects.

A spectral theory of magnetic-dipolar (magnetostatic) resonances in small ferrite particles does not presume the presence of the electric displacement current. The vectors $\vec{\mathcal{E}}$, $\vec{e}$, and $\vec{\mathfrak{D}}$, reflecting different aspects of the electric field in the *L*-mode spectrum, as well as the pseudo-electric field $\vec{\epsilon}$ for *G* modes, are polar vectors. At the same time, in frames of the magnetostatic-oscillation description (characterizing by negligibly small variation of electric energy in small magnetic objects with strong temporal dispersion of permeability at microwaves), these electric fields cannot by related to the electric-polarization effects both inside a ferrite and in abutting dielectrics outside a ferrite. So no transformations of the MDM spectra due to variation of dielectric parameters of a sample should be observed in experiments and numerical-simulation results. Nevertheless, recent microwave experiments [55] clearly show an influence of dielectrics on the MDM oscillations. In these experiments, the MDM spectrum transformation due to dielectric samples abutting to the surface of a ferrite disk has been demonstrated. As one can observe, such dielectric loadings result in transformations of the MDM-resonance peak positions with very small variations of the peak amplitudes. It was shown that for a higher permittivity of a dielectric loading one has bigger spectrum expansion. To explain the experiments with the MDM spectrum transformation, the *G*-mode analytical model has been used [55]. Based on this model, it was shown that the MDM spectrum is sensitive to the permittivity parameters of materials abutting to a ferrite disk due to the presence of the eigenelectric fluxes of the $\vec{\epsilon}$ fields. While an analysis in Ref. [55] explains the MDM spectrum expansion with a dielectric loading, it cannot give an answer why the entire MDM spectrum becomes shifted (with respect to frequency or a bias magnetic field) at such a loading. This shift of an entire MDM spectrum, well observed in our numerical and experimental studies [32], can be explained based on an analysis of the present paper.

From the *L*-mode solutions, one sees that (in neglect of the effective electric polarization originated from chiral-order magnetization in a ferrite) the electric fields $\vec{e}$ [see Eqs. (72), (73)] and the magnetic fields in classical magnetostatics problems [34] are completely dual with each other. Because of such a duality, we can assume the existence of the mechanical torque when the electric field $\vec{e}$ exerts on a test point electric dipole $\vec{p}^{(e)}$. This mechanical torque is defined as a cross product of the electric field $\vec{e}$ and the electric moment of the dipole:

$$\vec{\mathcal{N}}^{(e)} = \vec{p}^{(e)} \times \vec{e}. \qquad (100)$$

It makes the electric field $\vec{e}$ a physically observable quantity in a local point, both inside and outside a ferrite.

Let us suppose that a point electric dipole $\vec{p}^{(e)}$ is initially oriented along a disk axis. From the above analytical solutions for $L$ modes, as well as from the numerical-simulation results [17, 18, 22, 23], it follows that in a MDM ferrite disk there is a rotating electric field $\vec{\mathcal{E}}$. In a central region of a ferrite disk one has a homogeneous rotation of the $\vec{\mathcal{E}}$ field, which lies in a disk plane. Since there is no electric polarization effects due to the electric field $\vec{\mathcal{E}}$, an action of this field will result in precession of the electric dipole $\vec{p}^{(e)}$. The mechanical torque is equal to the time rate of change of angular momentum. As an initial assumption, we can suppose that the time rate of change of angular momentum is proportional to the time rate of change of orientation of the electric dipole $\vec{p}^{(e)}$. Based on Eq. (100), we can write:

$$\vec{\mathcal{N}}^{(e)} = \vec{p}^{(e)} \times \vec{\mathcal{E}} = -\frac{1}{\gamma^{(e)}} \frac{d\vec{p}^{(e)}}{dt}. \qquad (101)$$

where a quantity $-\frac{1}{\gamma^{(e)}}$ is a certain coefficient of proportionality. A phenomenological parameter $\gamma^{(e)}$ we formally introduced in an analogy with the gyromagnetic ratio $\gamma$ which relates the electron spin angular momentum and the electron magnetic moment [47, 48].

A ferrite is a dielectric material with a sufficiently big value of a dielectric permittivity ($\varepsilon_r = 15$). Suppose that we put a quasi-2D ferrite disk in an external homogeneous DC electric field $\vec{E}_0$ oriented along a disk axis. In this case, we will have the homogeneous electric polarization $\vec{p}^{(e)}$ inside a ferrite disk. A uniform array of identical dipoles oriented along a disk axis is equivalent to surface electric charges on disk planes which produce a depolarization field. Evidently, the DC electric field $\vec{E}_0$ (resulting in the constant electric polarization) does not cause any mechanical torque in the motion equation for polarization $\vec{p}^{(e)}$.

Suppose now that we excite the MDM oscillations in a ferrite disk. When a ferrite disk is placed in an external homogeneous DC electric field $\vec{E}_0$, the electric field $\vec{\mathcal{E}}$ of MDM oscillations will cause precession of electric polarization about the direction of a disk axis. For the electric polarization $\vec{p}^{(e)}$, one has the following precession equation:

$$\frac{d\vec{p}^{(e)}}{dt} = -\gamma^{(e)} \vec{p}^{(e)} \times \vec{\mathcal{E}}. \qquad (102)$$

Here we assume that at this precession there is a small deviation of vector $\vec{p}^{(e)}$ from the direction of vector $\vec{E}_0$. So, one can neglect variation of quantity of the electric polarization.

The torque exerting on the electric polarization $\vec{p}^{(e)}$ due to the electric field $\vec{\mathcal{E}}$ should be equal to reaction torque exerting on the magnetization $\vec{m}$ in a ferrite disk. In this reaction, however, one should take into account an "orbital" moment of the $\vec{\mathcal{E}}$ field. As we discussed above, in Section III, the existence of the power-flow-density vortices of the MDM oscillations presumes an angular momentum of the fields. So, the electric field $\vec{\mathcal{E}}$ of MDMs has both the "spin" and "orbital" angular momentums. It means that the electric polarization in a disk (in the presence of an external electric field) will be characterized with both the "spin" and "orbital" angular momentums. Due to the angular velocity of the "orbital" rotation of the electric polarization, the motion equation for magnetization will be modified. This modification, described by Eqs. (B5), (B6) Appendix B, will result in transformation of the MDM spectra.

When we put a dielectric loading above or (and) below a ferrite disk and apply to a structure a DC electric field oriented along a disk axis, we have two (or three) capacitances connected in series. The capacitance of a thin-film ferrite disk is much bigger than the capacitances of dielectric samples. So, surface electric charges on ferrite-disk planes will be mainly defined by the permittivity and geometry of dielectric samples. As a result, one will have MDM spectrum transformation dependable on parameters of the dielectric samples. In microwave experiments [32, 55], as well as in numerical analyses [17, 18, 22, 23, 32], we do not have external DC electric fields. In these studies, however, the electric polarization of a ferrite disk and dielectric samples takes place due to RF electric fields of electromagnetic waves propagating in a microwave waveguide. In such a case one has a rather more complicated process of an interaction of MDMs with polarized dielectrics. Nevertheless, the main physical aspects of this interaction discussed above for the DC electric polarization, will be applicable also in a case of the RF electric polarization. Further discussions on precession of electric polarization in a case of the RF electric fields and a role of this precession on transformation of the MDM spectra are given in Ref. [32].

With use of the microwave ME near-field structures one acquires an effective instrument for local characterization of topological properties of matter. Non-zero helicity density of the MDM near fields allows precise spectroscopic analysis of natural and artificial chiral structures at microwaves. This paves a way to creating pure microwave devices for separation of biological and drug enantiomers. This also may give an answer to a controversial issue whether or not microwave irradiation can exert a non thermal effect on biomolecules [102]. In a view of these discussions, it is worth noting that in biological structures, microwave radiation can excite certain rotational transitions and some extraordinary effects, which cannot be explained as classical heating effects, give the clearest examples of a possible specific action created by a microwave radiation field [102].

*C. On magnetoelectric interactions in artificial electromagnetic materials: Do really magnetoelectric interactions exist in bianisotropic metamaterials.*

A general idea of magnetoelectric (ME) metamaterials is to obtain coupling between the electric and magnetic fields separately from such a coupling in Maxwell's equations. So called bianisotropic metamaterials were conceived as such a kind of ME metamaterials, where one supposes existence of local cross-polarization terms together with the electric- and magnetic-polarization terms. However, the shown peculiar properties of ME fields give us the possibilities for some critical analysis of near-field "ME interactions" in such artificial electromagnetic materials as bianisotropic metamaterials.

In a case of a metallic bianisotropic particle [such, for example, as a split-ring resonator (SRR) or an omega-particle], an incident EM wave experiences phase shift between the electric and magnetic fields. Every bianisoropic particle (BAP) behaves like a small electromagnetic antenna. In dilute structures, an interaction between BAPs is due to *electromagnetic radiation*. In dense metamaterials, quasistatic interactions between BAPs are the following: there are only quasielectrostatic (between the capacitance parts) and quasimagnetostatic (between the inductive parts) interactions. There are no quasistatic ME interactions, since no internal ME energy are presumed in such elements. In a dense BAP lattice, there are, in fact, only the electric and magnetic interaction constants ($C_E$ and $C_M$), but there are no ME-interaction constants (no $C_{ME}$). The only way for ME coupling is via the radiation (retardation) effects. So bianisotropy in these metamaterials is possible only due to effects of nonlocality. There are non-standing-wave currents inside a split-ring resonator or an omega-particle. We cannot speak about the *microscopic* electrodynamics of bianisotropic metamaterials, since there are no internal motion processes associated with microscopic ME fields. One can adduce also other argumentation. Let us consider a small object. There is an object with sizes much less than a free-space wavelength. We know that inside this object one has transformation of electric energy to magnetic energy and vice versa. The object is an electromagnetic *LC* oscillator. Since there is an open structure, we can use point electric and magnetic probes (placed in definite positions of the object) to see the electric-to-magnetic and magnetic-to-electric energy transformations. Suppose now that an electromagnetic wave incident on our object. Can the wave scattered from this particle bear the imprint of the *LC*-oscillation process of the electric-to-magnetic and magnetic-to-electric energy transformations occurring inside the object? Certainly, can. But on sizes comparable, to some extent, with the free-space wavelength, since only on such a scale there are energy transformations in electromagnetic waves propagating in vacuum. There is an evident reason for this. In classical electrodynamics, one does not have locally coupled electric-plus-magnetic sources [34]. An array of such *LC* particles is just a classical diffractional structure with specific field polarization effects. Let us consider now a small open structure with the electric- and magnetic-field oscillations, but not the *LC* oscillator. It means that we cannot separate definitely the regions of the electric and magnetic fields. There is a particle with specific intrinsic dynamical process in the material. The near-field region of such a particle should be characterized by a specific field structure since the electric- and magnetic-field components of the near fields are originated from intrinsic dynamical process in material of the object. They are not coupled via Maxwell equations. We call such near fields as the magnetoelectric fields. EM fields scattered from the ME-field particles should have a topological structure with the presence of geometrical phases.

In a classical consideration, ME fields originated from a lossless MDM ferrite particle are potential near fields. In the near-field region one has the Laplace equation for magnetostatic potential $\psi$ ($\nabla^2\psi = 0$) as well as the Laplace equation for electrostatic potential $\Phi^{(e)}$ ($\nabla^2\Phi^{(e)} = 0$). At the MDM resonances, the potentials $\psi$ and $\Phi^{(e)}$ are coupled due to magnetization processes inside a ferrite disk particle. The MDM magnetization processes (described phenomenologically by the Landau-Lifshitz equation) are related to electron mass via gyromagnetic ratio [48]. Since mass is a measure of inertia, one has causality in ME dynamical responses of a ferrite particle. A composite based on ferrite MDM particles will behave as a causal ME metamaterial. In a case of lossless metallic "bianisotropic particles" such, for example, as SRRs, a character of a dynamical response is completely different. A "magnetic part" of a SRR is characterized by inertia due to inductance (inductance is like mechanical inertia). If you have a wire in which you try to change the current, that change in current produces an electric field back upon the wire which tries to oppose that increase in current. The field opposing the change does not travel at infinite velocity so it always lags a bit behind the potential driving the current. For a small metallic ring this lag is not due to non-locality. At the same time, the "electric part" (in neglect of any plasmon resonances and when a SRR is at rest) has no inertia in local dynamic response. The "ME" properties of a SRR particle are due to retardation effects. The fields surrounding such particles are described by the Helmholtz equation. One does not have the Laplace equations for electric and magnetic potentials and no coupling between such potentials.

In a case of ferrite MDM disks we have a localized ME field. There is a "microscopic" field entity. We can create a ME lattice based on or *G*-mode interacting ME particles, or *L*-mode interacting ME particles, or combined *L*- and *G*-mode interacting ME particles. In these structures, electric fluxes of one ME particle interacts with magnetization dynamics of another ME particle. Also magnetic fluxes of one ME particle interacts with magnetization dynamics of another ME particle. It means that in the interaction process (between ME particle) both the electric and magnetic components of the ME field take part. One cannot separate the regions of "pure" electric and "pure" magnetic fields. Here we stress on the following. Subwavelength interaction between particles with ME properties – ME particles – can be realized only by "microscopic" ME fields. No such an interaction is possible via electromagnetic near fields.

**VII. Conclusion**

Interaction between electromagnetic waves and matter on a subwavelength scale opens a new field of studies: near-field electrodynamics. In the near-field electrodynamics, space and time can be coupled in a manner different from the far-field electrodynamics. This may create a new type of the field substance. In this paper, we showed that in a close proximity of a MDM ferrite disk there exists a quantized near field which is characterized by peculiar symmetry properties. This is a topological, curved space-time field. Such an entity, differing from the known electromagnetic near-field structures, we call a magnetoelectric field. A near field of the MDM particle – the ME field – is a quasimagnetostatic field. This field, however, cannot be considered as a field dual to known quasielectrostatic fields. We showed that there is a fundamental difference between the observed ME fields and the near fields originated from plasmon-oscillation particles. One of important distinctive features of the ME fields is the presence of the helicty structure in a vacuum near-field region.

The main properties of the ME fields become clear when one analyses spectral solutions for the MS-potential wave function $\psi$ in a ferrite-disk particle. To make the MDM spectral problem analytically integrable, two approaches were suggested. These approaches, distinguishing by differential operators and boundary conditions used for solving the spectral problem, give two types of the MDM oscillation spectra in a quasi-2D ferrite disk: The *G*- and *L*-mode spectra. The MS-potential wave function $\psi$ manifests itself in different manners for every of these types of spectra. In this paper, we studied the field structures for the *G*- and *L*-mode spectra. We also analyzed possible interactions between these two-type spectral solutions. Based on the spectral analysis, we showed that for ME fields originated from MDM ferrite particles one can observe a torsion degree of freedom.

We discussed the mechanisms of interaction between microwave ME fields and dielectric samples. We propose also that with use of the microwave ME near-field structures one may acquire an effective instrument for local characterization of special topological properties of matter. This, in particular, will allow realization of microwave devices for precise spectroscopic analysis of materials with chiral structures such, for example, as biological and drug enantiomers. In a view of physical properties of the ME fields, we discussed the mechanisms of near-field coupling between ME particles. We stress on the fact that the subwavelength interaction between particles with ME properties can be realized only by "microscopic" ME fields and that no such an interaction is possible in so called bianisotropic metamaterials.

**Appendix A: Magnetic-current vector potentials in MS oscillations**

In classical electromagnetism, the vector potential $\vec{A}$ is introduced for convenience in solving magnetostatic problems with use of the representation for the magnetic flux density as $\vec{B} = \nabla \times \vec{A}$. Frequently, the term vector potential is referred as the magnetic vector potential. For the magnetostatic problems, the relation of the vector potential $\vec{A}$ to the electric-current source $\vec{j}^{(e)}$ is [34]:

$$\nabla^2 \vec{A} = -\frac{4\pi}{c} \vec{j}^{(e)}. \qquad (A1)$$

When the fields are time-varying, one can introduce auxiliary potential functions and specify the $\vec{E}$ and $\vec{B}$ fields as

$$\vec{E} = -\frac{1}{c}\frac{\partial \vec{A}}{\partial t} - \nabla \Phi; \qquad \vec{B} = \nabla \times \vec{A}, \qquad (A2)$$

where $\Phi$ is the scalar potential function and $\vec{A}$ is the vector potential function. In a case of the time-varying fields, the relation of the vector potential function $\vec{A}$ to the electric-current source $\vec{j}^{(e)}$ is [34]:

$$\nabla^2 \vec{A} - \frac{1}{c^2}\frac{\partial^2 \vec{A}}{\partial t^2} = -\frac{4\pi}{c} \vec{j}^{(e)}. \qquad (A3)$$

Because of the electric-current sources for the vector potential, both for the magnetostatic and time-varying fields, one can also call the vector potential $\vec{A}$, as the electric-current vector potential. To stress on this definition, we rewrite Eq. (A3) as

$$\nabla^2 \vec{A}^{(e)} - \frac{1}{c^2}\frac{\partial^2 \vec{A}^{(e)}}{\partial t^2} = -\frac{4\pi}{c} \vec{j}^{(e)}, \qquad (A4)$$

where $\vec{A}^{(e)}$ means the electric-current vector potential.

In spite of the fact that no magnetic charges and no motion equations for magnetic charges are known in nature, because of the electromagnetic duality one can formally introduce magnetic currents in Maxwell equations. This formal procedure allows solving numerous electrodynamics problems [78, 79]. For the electrodynamic vector potential caused by a magnetic current $\vec{j}^{(m)}$ we have the wave equation [78, 80]:

$$\nabla^2 \vec{A}^{(m)} - \frac{1}{c^2}\frac{\partial^2 \vec{A}^{(m)}}{\partial t^2} = -\frac{4\pi}{c} \vec{j}^{(m)}, \qquad (A5)$$

where $\vec{A}^{(m)}$ we call the magnetic-current vector potential. The Eq. (A5) was obtained when one uses the following representation for the electric displacement field

$$\vec{D} = \nabla \times \vec{A}^{(m)}. \qquad (A6)$$

With such a representation, one has from the Maxwell equations for the magnetic field:

$$\vec{H} = -\frac{1}{c}\frac{\partial \vec{A}^{(m)}}{\partial t} - \nabla \psi, \qquad (A7)$$

where $\psi$ is the magnetic scalar potential. It is worth noting that, referring to representation (A6), in some publications, the magnetic-current vector potential $\vec{A}^{(m)}$ is called as the electric vector potential. There is, for examples, the study of the fields of toroidal solenoids [81]. At the same time, such terms as the electric-current vector potential (regarding the vector $\vec{A}^{(e)}$ caused by an electric current) and the magnetic-current vector potential (regarding the vector $\vec{A}^{(m)}$ caused by a magnetic current), used also in our study, one can find in Ref. [82].

As we discussed above, in a case of MS oscillations there are no effects of the electromagnetic retardation. The term $\dfrac{\partial \vec{A}^{(m)}}{\partial t}$ does not define the magnetic field $\vec{H}$ (in a case of the MS oscillations the magnetic field is $\vec{H} = -\nabla \psi$) and thus Eq. (A5) is written as

$$\nabla^2 \vec{A}^{(m)} = -\frac{4\pi}{c} \vec{j}^{(m)}. \tag{A8}$$

There is a dual situation with respect to the problem described by Eq. (A1). Such an equation one obtains immediately from Eqs. (37) and (38) and taking into account a proper gauge (the Coulomb gauge [34]) transformation.

Based on Eqs. (37) and (38), we have

$$\nabla^2 \vec{A}_\epsilon^{(m)} - \nabla \left( \nabla \cdot \vec{A}_\epsilon^{(m)} \right) + \frac{4\pi}{c} \vec{j}^{(m)} = 0. \tag{A9}$$

This equation shows that formally two types of gauges are possible. In the first type of a gauge we have:

$$\nabla \cdot \vec{A}_\epsilon^{(m)} = 0 \tag{A10}$$

and, therefore,

$$\nabla^2 \vec{A}_\epsilon^{(m)} = -\frac{4\pi}{c} \vec{j}^{(m)}. \tag{A11}$$

The second type of a gauge is written as

$$\nabla \left( \nabla \cdot \vec{A}_\epsilon^{(m)} \right) - \frac{4\pi}{c} \vec{j}^{(m)} = 0 \tag{A12}$$

and, therefore,

$$\nabla^2 \vec{A}_\epsilon^{(m)} = 0. \tag{A13}$$

The last equation shows that any sources of the electric field are not defined and thus the electric field is not defined at all. So only the first type of a gauge, described by Eq. (A10) and resulting in Eq. (A11), should be taken into account.

**Appendix B: Transformation of a magnetic susceptibility tensor to a rotating reference frame**

Let $S'(x', y', z')$ be a frame of reference rotating with respect to the laboratory frame $S(x, y, z)$ with an angular velocity represented by a vector $\vec{\omega}_{frame}$. The space-time transformation from the laboratory frame $S$ to the rotating frame $S'$ is given by (see e. g. [87])

$$x' = x\cos(\omega_{frame} t) + y\sin(\omega_{frame} t),$$

$$y' = -x\sin(\omega_{frame} t) + y\cos(\omega_{frame} t), \quad (B1)$$

$$z' = z, \qquad t' = t.$$

It is evident that the vector differential operator $\vec{\nabla}$ is invariant under the transformation (B1):

$$\vec{\nabla}' = \vec{\nabla}. \quad (B2)$$

According to the general law of relative motion, the time derivative of any time-dependent vector $\vec{C}(t)$, computed in the laboratory frame $S$, and the time derivative computed in the rotating frame $S'$, are related through (see e.g. [88])

$$\left(\frac{d\vec{C}}{dt}\right)_{lab} = \left(\frac{d\vec{C}}{dt}\right)_{rot} + \vec{\omega}_{frame} \times \vec{C}. \quad (B3)$$

The motion of the magnetic moment in the laboratory frame is described by the equation

$$\frac{d\vec{m}}{dt} = -\gamma\, \vec{m} \times \vec{H}. \quad (B4)$$

Based on Eq. (B1), we have the motion equation of the magnetic moment in the rotating frame [89]

$$\left(\frac{d\vec{m}}{dt}\right)_{rot} = -\gamma\, \vec{m} \times \left(\vec{H} - \frac{\vec{\omega}_{frame}}{\gamma}\right). \quad (B5)$$

For the motion of the magnetic moment in the rotating frame we have an effective field

$$\vec{H}_{eff} = \vec{H} - \frac{\vec{\omega}_{frame}}{\gamma}, \quad (B6)$$

which is the sum of the laboratory-frame field $\vec{H}$ and a fictitious field $-\frac{\vec{\omega}_{frame}}{\gamma}$.

Now consider a ferromagnet. Because of the presence of the exchange interaction between the spins of the individual atoms, the magnetic moment of the ferromagnet may be regarded as rigid provided only the temperature of the ferromagnet is sufficiently small. For the magnetic moment density of the ferromagnet, $\vec{M}$, we have the motion equation in the rotating frame similar to Eq. (B5):

$$\left(\frac{d\vec{M}}{dt}\right)_{rot} = -\gamma \, \vec{M} \times \left(\vec{H} - \frac{\vec{\omega}_{frame}}{\gamma}\right). \tag{B7}$$

We represent the total field $\vec{H}$ and the magnetization $\vec{M}$ in the rotating frame as sums of the DC and RF components

$$\vec{H} = \vec{H}_0 + \vec{H}_\sim, \qquad \vec{M} = \vec{M}_0 + \vec{m}_{rot} \tag{B8}$$

and suppose that

$$\vec{H}_\sim \ll \vec{H}_0, \qquad \vec{m}_{rot} \ll \vec{M}_0. \tag{B9}$$

In this case, Eq. (B7) acquires the form:

$$\left(\frac{d\vec{m}}{dt}\right)_{rot} + \gamma \, \vec{m}_{rot} \times \left(\vec{H}_0 - \frac{\vec{\omega}_{frame}}{\gamma}\right) = -\gamma \, \vec{M}_0 \times \vec{H}_\sim. \tag{B10}$$

Let $\vec{H}_\sim$ and $\vec{m}_{rot}$ be the time harmonic functions characterized by frequency $\omega$ ($\sim e^{i\omega t}$). For complex amplitudes in the rotating frame, Eq. (B10) is rewritten as

$$i\omega \, \vec{m}_{rot} + \gamma \, \vec{m}_{rot} \times \left(\vec{H}_0 - \frac{\vec{\omega}_{frame}}{\gamma}\right) = -\gamma \, \vec{M}_0 \times \vec{H}_\sim. \tag{B11}$$

We consider the case when the vectors $\vec{H}_0$ and $\vec{M}_0$ are directed along the $z$ axis and when the vector $\vec{\omega}_{frame}$ is oriented along the $z$ axis as well. From Eq. (B11), one has the following equations in Cartesian coordinates:

$$i\omega \, (m_x)_{rot} + (\gamma H_0 - \omega_{frame})(m_y)_{rot} = \gamma \, M_0 (H_\sim)_y$$

$$-(\gamma H_0 - \omega_{frame})(m_x)_{rot} + i\omega \, (m_y)_{rot} = -\gamma \, M_0 (H_\sim)_x \tag{B12}$$

$$(m_z)_{rot} = 0$$

This gives the following equation:

$$\vec{m}_{rot} = \ddot{\chi}_{rot} \cdot \vec{H}_\sim, \tag{B13}$$

where

$$\ddot{\chi}_{rot} = \begin{pmatrix} \chi_{rot} & i(\chi_a)_{rot} & 0 \\ -i(\chi_a)_{rot} & \chi_{rot} & 0 \\ 0 & 0 & 0 \end{pmatrix} \tag{B14}$$

and

$$\chi_{rot} = \frac{\gamma M_0 (\omega_H - \omega_{frame})}{(\omega_H - \omega_{frame})^2 - \omega^2}, \qquad (\chi_a)_{rot} = \frac{\gamma M_0 \omega}{(\omega_H - \omega_{frame})^2 - \omega^2}, \qquad \omega_H \equiv \gamma H_0. \qquad (B15)$$

We have to note here that the frequency $\omega_{frame}$ is a positive quantity. In a particular case when $\omega_{frame} = \omega_H$, we have $\chi_{rot} = 0$ and $(\chi_a)_{rot} = -\frac{\gamma M_0}{\omega}$.

In the rotating frame, one has for magnetic flux density

$$\vec{B}_{rot} = \ddot{\mu}_{rot} \cdot \vec{H}_\sim, \qquad (B16)$$

where the magnetic permeability tensor is defined as

$$\ddot{\mu}_{rot} = \ddot{I} + 4\pi \ddot{\chi}_{rot}. \qquad (B17)$$

Here $\ddot{I}$ is unit matrix.

We consider now the frequencies corresponding to the MDM resonances in a ferrite disk. At these resonances, one has the rotating magnetic fields. For $\omega = \omega_{frame} = \omega_{res}$, we obtain

$$(\chi_{rot})_{res} = \frac{\gamma M_0 (\omega_H - \omega_{res})}{(\omega_H)^2 - 2\omega_H \omega_{res}}, \qquad [(\chi_a)_{rot}]_{res} = \frac{\gamma M_0 \omega_{res}}{(\omega_H)^2 - 2\omega_H \omega_{res}}. \qquad (B18)$$

For diagonal and off-diagonal components of the permeability tensor, we have

$$(\mu_{rot})_{res} = \frac{\omega_H (\omega_H - 2\omega_{res} + \omega_M) - \omega_M \omega_{res}}{(\omega_H)^2 - 2\omega_H \omega_{res}}, \qquad [(\mu_a)_{rot}]_{res} = \frac{\omega_M \omega_{res}}{(\omega_H)^2 - 2\omega_H \omega_{res}}, \qquad (B19)$$

where $\omega_M \equiv \gamma 4\pi M_0$. When $\omega_{res} = \omega_{frame} > \omega_H$, there are $(\chi_{rot})_{res} > 0$ and $[(\chi_a)_{rot}]_{res} < 0$ resulting in $(\mu_{rot})_{res} > 0$ and $[(\mu_a)_{rot}]_{res} < 0$. In a case when $\omega_{res} = \omega_{frame} < \omega_H$, one has $(\chi_{rot})_{res} < 0$ and $[(\chi_a)_{rot}]_{res} > 0$. This gives $(\mu_{rot})_{res} < 0$ and $[(\mu_a)_{rot}]_{res} > 0$.

**Figure captions**

Fig. 1. Homogeneous rotation of the electric field $\vec{e}$. A bias magnetic field $\vec{H}_0$ is directed toward the observer normally to a disk plane. When (for a given radius and a certain time phase $\omega t$) an azimuth angle $\theta$ varies from 0 to $2\pi$, the electric-field vector accomplishes the geometric-phase rotation.

Fig. 2. Qualitative distributions of the helicity density $F$ for the fields above $\left(z \geq \dfrac{d}{2}\right)$ and below $\left(z \leq -\dfrac{d}{2}\right)$ a ferrite disk for different orientations of a bias magnetic field. The unit vector $\vec{e}_z$ is oriented along z-axis.

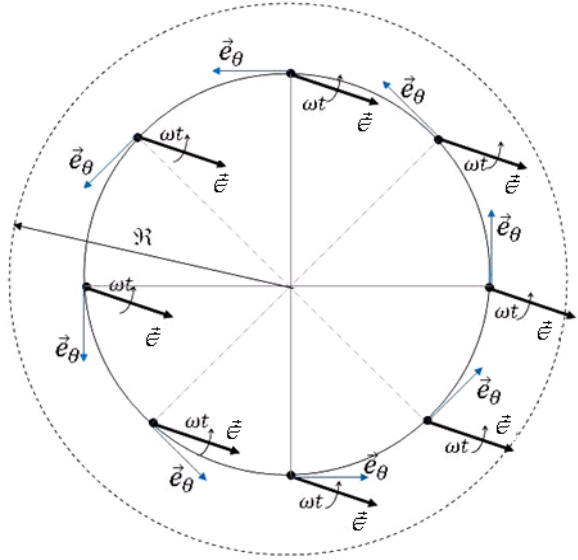

Fig. 1. Homogeneous rotation of the electric field $\vec{E}$. A bias magnetic field $\vec{H}_0$ is directed toward the observer normally to a disk plane. When (for a given radius and a certain time phase $\omega t$) an azimuth angle $\theta$ varies from 0 to $2\pi$, the electric-field vector accomplishes the geometric-phase rotation.

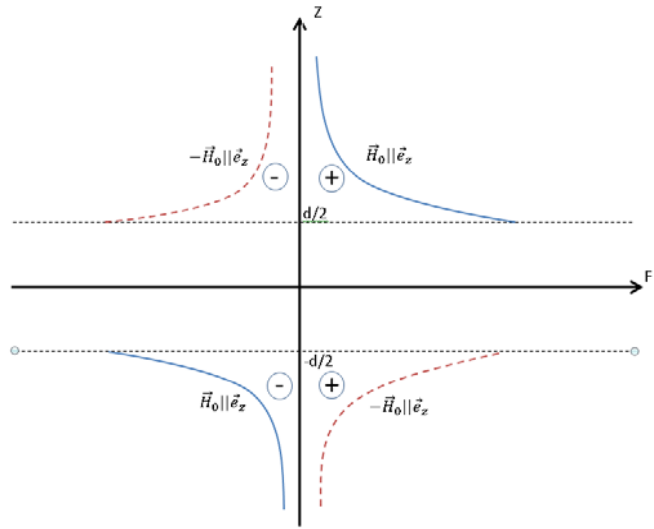

Fig. 2. Qualitative distributions of the helicity density $F$ for the fields above $\left(z \geq \dfrac{d}{2}\right)$ and below $\left(z \leq -\dfrac{d}{2}\right)$ a ferrite disk for different orientations of a bias magnetic field. The unit vector $\vec{e}_z$ is oriented along $z$-axis.